\documentclass[10pt,conference]{IEEEtran}
\usepackage{cite}
\usepackage{amsmath,amssymb,amsfonts}
\usepackage{algorithmic}
\usepackage{graphicx}
\usepackage{textcomp}
\usepackage{xcolor}
\usepackage[hyphens]{url}
\usepackage{fancyhdr}
\usepackage{hyperref}

\usepackage{braket}
\usepackage{subcaption}
\usepackage{overpic}
\usepackage{multirow}

\usepackage{multicol}

\newcommand{\hpcayear}{2025}

\newcommand{\figvspaceelim}{-2mm}

\newcommand{\hpcasubmissionnumber}{1057}
\title{LSQCA: Resource-Efficient Load/Store Architecture for Limited-Scale Fault-Tolerant Quantum Computing}
\usepackage{tikz}
\usepackage{textcomp}
\usepackage{hyperref}
\usepackage{lipsum}

\newcommand\copyrighttext{%
  \footnotesize \textcopyright 2025 IEEE. Personal use of this material is permitted.  Permission from IEEE must be obtained for all other uses, in any current or future media, including reprinting/republishing this material for advertising or promotional purposes, creating new collective works, for resale or redistribution to servers or lists, or reuse of any copyrighted component of this work in other works.
}
\newcommand\copyrightnotice{%
\begin{tikzpicture}[remember picture,overlay]
\node[anchor=south,yshift=10pt] at (current page.south) {\fbox{\parbox{\dimexpr\textwidth-\fboxsep-\fboxrule\relax}{\copyrighttext}}};
\end{tikzpicture}%
}
\def\hpcacameraready{} % Uncomment to build camera-ready version

\newcommand\hpcaauthors{Takumi Kobori$\dagger$, Yasunari Suzuki$\ddagger$, Yosuke Ueno*, Teruo Tanimoto**, Synge Todo$\dagger$, and Yuuki Tokunaga$\ddagger$}
\newcommand\hpcaaffiliation{The University of Tokyo$\dagger$, NTT Corporation$\ddagger$, RIKEN*, Kyushu University**}
\newcommand\hpcaemail{takumi.kobori@phys.s.u-tokyo.ac.jp, yasunari.suzuki@ntt.com, yosuke.ueno@riken.jp,\\ tteruo@kyudai.jp, wistaria@phys.s.u-tokyo.ac.jp, yuuki.tokunaga@ntt.com}

\author{
  \ifdefined\hpcacameraready
    \IEEEauthorblockN{\hpcaauthors{}}
      \IEEEauthorblockA{
        \hpcaaffiliation{} \\
        \hpcaemail{}
      }
  \else
    \IEEEauthorblockN{\normalsize{HPCA \hpcayear{} Submission
      \textbf{\#\hpcasubmissionnumber{}}} \\
      \IEEEauthorblockA{
        Confidential Draft \\
        Do NOT Distribute!!
      }
    }
  \fi 
}

\fancypagestyle{camerareadyfirstpage}{%
  \fancyhead{}
  
  \fancyhead[C]{
    \ifdefined\aeopen
    \parbox[][12mm][t]{13.5cm}{\hpcayear{} IEEE International Symposium on High-Performance Computer Architecture (HPCA)}    
    \else
      \ifdefined\aereviewed
      \parbox[][12mm][t]{13.5cm}{\hpcayear{} IEEE International Symposium on High-Performance Computer Architecture (HPCA)}
      \else
      \ifdefined\aereproduced
      \parbox[][12mm][t]{13.5cm}{\hpcayear{} IEEE International Symposium on High-Performance Computer Architecture (HPCA)}
      \else
      \parbox[][0mm][t]{13.5cm}{\hpcayear{} IEEE International Symposium on High-Performance Computer Architecture (HPCA)}
    \fi 
    \fi 
    \fi 
    \ifdefined\aeopen 
      \includegraphics[width=12mm,height=12mm]{ae-badges/open-research-objects.pdf}
    \fi 
    \ifdefined\aereviewed
      \includegraphics[width=12mm,height=12mm]{ae-badges/research-objects-reviewed.pdf}
    \fi 
    \ifdefined\aereproduced
      \includegraphics[width=12mm,height=12mm]{ae-badges/results-reproduced.pdf}
    \fi
  }
  \fancyfoot[C]{}
}
\begin{document}
\maketitle

%Enables the camera ready header and footer
\ifdefined\hpcacameraready 
  \thispagestyle{camerareadyfirstpage}
  \pagestyle{empty}
\else
  \thispagestyle{plain}
  \pagestyle{plain}
\fi

\newcommand{\hpcaheight}{0mm}
\ifdefined\eaopen
\renewcommand{\hpcaheight}{12mm}
\fi

%%%%%%%%%%%%%%%%%%%%%%%%%%%%%%%%%%%%%%%%
%%%%%%%% -- PAPER CONTENT STARTS -- %%%%%%%%%
\copyrightnotice

\begin{abstract}

Current fault-tolerant quantum computer (FTQC) architectures utilize several encoding techniques to enable reliable logical operations with restricted qubit connectivity. However, such logical operations demand additional memory overhead to ensure fault tolerance. Since the main obstacle to practical quantum computing is the limited qubit count, our primary mission is to design floorplans that can reduce memory overhead without compromising computational capability. Despite extensive efforts to explore FTQC architectures, even the current state-of-the-art floorplan strategy devotes 50{\%} of memory space to this overhead, not to data storage, to guarantee unit-time random access to all logical qubits. 

In this paper, we propose an FTQC architecture based on a novel floorplan strategy, Load/Store Quantum Computer Architecture (LSQCA), which can achieve almost 100{\%} memory density. The idea behind our architecture is to separate the whole memory regions into small computational space called Computational Registers (CR) and space-efficient memory space called Scan-Access Memory (SAM). We define an instruction set for these abstract structures and provide concrete designs named point-SAM and line-SAM architectures. With this design, we can improve the memory density by allowing variable-latency memory access while concealing the latency with other bottlenecks. We also propose optimization techniques to exploit properties of quantum programs observed in our static analysis, such as access locality in memory reference timestamps.
Our numerical results indicate that LSQCA successfully leverages this idea. In a resource-restricted situation, a specific benchmark shows that we can achieve approximately 90{\%} memory density with 5{\%} increase in the execution time compared to a conventional floorplan, which achieves at most 50{\%} memory density for unit-time random access. Our design is defined as an abstract form, making this principle ubiquitous and applicable to a wide range of quantum devices, qubit-connectivity configurations, and error-correcting codes.

\end{abstract}

\section{Introduction}\label{sec:introduction}
Fault-tolerant quantum computing (FTQC) is expected to enable faster computation that surpasses classical computers~\cite{nielsen2002quantum, shor1999polynomial, harrow2009quantum, low2019hamiltonian,gidney2021factor, ur2023quantum, bauer2020quantum}. 
The major problems in realizing FTQCs are their high error rates and the limited number of available qubits.
Quantum error correction~(QEC)~\cite{Shor1995Sscheme,Steane1996} is a promising solution to reduce error rates by protecting qubits with error-correcting codes.
Among the various types of quantum error-correcting codes, surface codes~\cite{kitaev2003quantum,bravyi1998quantum,dennis2002topological} have gained attention due to their high performance and practical implementability. Thus, most FTQC proposals are based on this design~\cite{Krinner2022,sivak2023real,google2023suppressing,bluvstein2024logical,zhang2024demonstrating}, and the optimization of quantum algorithms and programs is dedicated to surface-code-based architectures~\cite{horsman2012surface,fowler2018low,litinski2019game,yoshioka2023hunting,gidney2021factor,joshua2022reliably}. 
A drawback of using surface-code-based architecture is the memory overhead required to perform logical operations on the surface-code qubits. Standard logical operations such as lattice surgery and code deformation demand temporal allocations of spatially neighboring surface-code cells as an auxiliary space, which makes the problem of limited qubit counts more serious.
Therefore, a strategy to find high-memory-density floorplans for data logical-qubit cells and auxiliary cells is demanded to execute large quantum programs with a small number of qubits, and massive efforts have been paid to find efficient floorplan strategies~\cite{beverland2022surface,chamberland2022universal,beverland2022assessing,lee2021even}.

However, the existing floorplans have the problem of the resources required for operations increasing proportionally with the number of data cells. Due to this problem, even the state-of-the-art floorplan suffers from low memory density, which can use at most 50\% or 67\% memory space~\cite{beverland2022assessing,lee2021even}.
This difficulty arises because they aim to maintain parallelism of logical operations as much as possible, ensuring that all logical qubits are immediately ready for operations. 
From a computer science perspective, this approach resembles a system where all the memory cells are composed of registers, which can be inefficient, especially in a resource-restricted scenario. 
Additionally, considering the overall computation in FTQC, various potential bottlenecks exist beyond the parallel execution of operations. For example, producing magic states with small error rates is necessary for universal FTQC but is costly in terms of both time and space. Thus, the magic-state generation would be one of the dominant bottleneck factors in resource-restricted FTQC~\cite{litinski2019game,bravyi2005universal,bravyi2012magic,fowler2013surface}.
Therefore, there is a strong demand for designing a floorplan strategy that can reduce the required qubit count with a small penalty of operational parallelism while concealing the overhead of execution time by other bottlenecks.

\begin{figure}[t]
    \centering
    \includegraphics[width=0.5\textwidth]{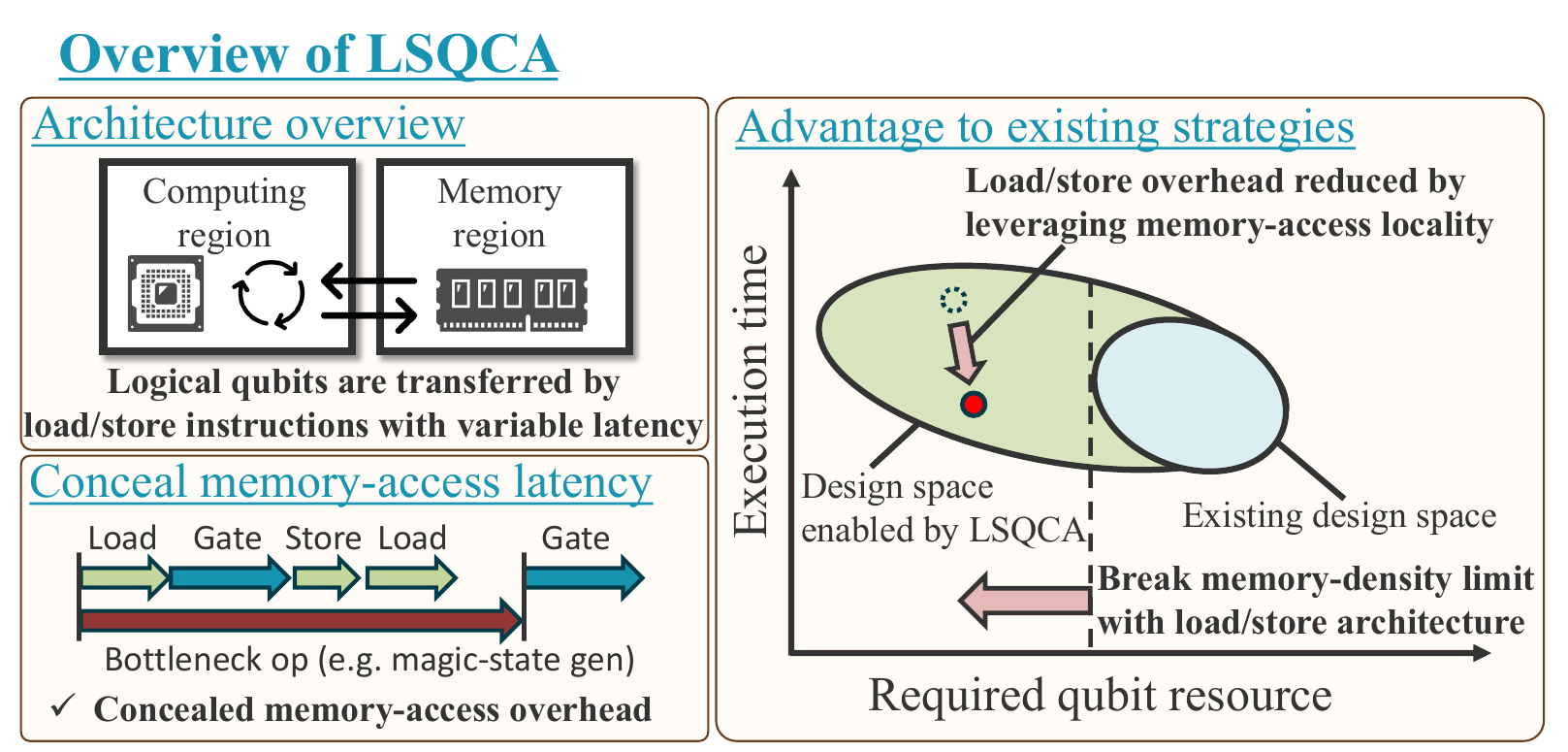}
    \caption{The positioning of our proposal, LSQCA, which breaks the memory density limit of conventional floorplans and enables the exploration of the area in the trade-off between required resources and execution time where conventional floorplans cannot reach.}
    \label{fig:research_position}
    \vspace{\figvspaceelim}
\end{figure}

In this paper, we propose an FTQC architecture based on a novel floorplan strategy called Load/Store Quantum Computer Architecture (LSQCA). This architecture is designed to combine a small computational region and space-efficient memory space. To execute various application programs in this architecture, we propose an instruction set architecture supporting data movement between the two regions as an abstraction, referring to the relation between the register and memory in conventional computers.
For surface-code-based FTQC, we proposed a practical design of LSQCA that consists of Computational Register (CR) as a computational region and Scan-Access Memory (SAM) as a memory space.
In this architecture, logical qubits can be loaded from SAM to CR, and logical operations can be performed in CR. Then, the loaded qubits are stored back in SAM. The SAM can achieve nearly 100{\%} memory density by allowing variable-latency memory access while suppressing the increase of average latency by concealing it with other bottlenecks~(Fig.\,\ref{fig:research_position}). The design with CR and SAM also allows flexible tuning of the trade-off between memory overhead and access latency.

LSQCA can be considered a form of memory hierarchy, which can improve average latency by utilizing the access locality in quantum programs.
Conventional computers leverage the spatial and temporal access locality in programs by the hierarchical memory subsystem~\cite{patterson2016computer}. This strategy ensures the scalability of memory capacity while suppressing average access latency.
We performed static analysis on memory reference patterns for FTQC algorithms and found that they have access locality.
Then, we integrate locality-aware data movement strategies, and our numerical calculations show that the LSQCA can exploit the locality. We also introduce several crucial concepts to optimize LSQCA, such as multi-bank memories, in-memory operations, and hybrid floorplans.

Our contributions are as follows:
\begin{itemize}
    \item We propose a novel architecture named LSQCA, introducing the concept of register-memory data movements to FTQC floorplan designs.
    \item We propose an instruction set architecture of LSQCA and provide designs with CR and SAM. We provide two types of designs compatible with surface-code-based FTQC: point-SAM and line-SAM architectures, exhibiting different characteristics in terms of memory efficiency, latency, and the utilization of access localities.
    \item The performance and the execution time overhead of LSQCA are evaluated by numerical simulations with realistic quantum programs and various configurations. The results show that LSQCA achieves higher memory density than conventional floorplans with a negligible or modest increase in execution time.
    \item Under a resource-restricted scenario, LSQCA can achieve 87{\%} and 92{\%} memory density at the cost of 6{\%} and 7{\%} increase in execution time for multiplier and SELECT circuits compared to conventional floorplans, respectively, while conventional floorplans demand 50\% of memory space to guarantee unit-time access to arbitrary logical qubits.
\end{itemize}
Since our architecture is defined as an abstracted form, this principle can be applied to a wide range of qubit devices, connectivity configurations, logical operations methods, and error-correcting codes.
In future work, we expect that a more sophisticated instruction scheduler to handle variable memory-access latencies introduced by the LSQCA can further minimize the memory access overhead. 
For example, variable memory access latencies might be efficiently concealed by leveraging techniques commonly employed in conventional computing, such as strategic data allocation or prefetching based on access pattern detection.

\section{Background}\label{sec:background}
This section explains the basic part of FTQC. Due to the page limit, we focus on crucial points in our paper. For details on basic theories, please refer to Refs.\,~\cite{nielsen2002quantum,fowler2012towards,fowler2018low,litinski2019game}.

\subsection{Advantage of Quantum Computing}
Quantum computing is expected to process several scientific problems exponentially faster than those based on classical mechanics~\cite{nielsen2002quantum}. 
There are several difficulties in developing quantum computers. The main issue is the high error rate of qubits. This problem can be solved by QEC techniques described later, but QEC imposes significant overhead on qubit count and latency. A typical QEC strategy demands hundreds of qubits for constructing a single reliable logical qubit, and the latency for logical operation takes a few tens of microseconds. Additionally, the available number of qubits is limited due to the difficulties in integrating quantum devices. This situation motivates us to find application domains where we can easily demonstrate the advantage of quantum computing and establish compilation methods for them with a few qubits. Currently, solving integer factoring~\cite{gidney2021factor} and analyzing the spectrum of condensed matter or molecules~\cite{yoshioka2023hunting,babbush2018encoding} are considered promising targets.

\subsection{Surface Codes}
\begin{figure}[t]
    \centering \includegraphics[width=0.5\textwidth]{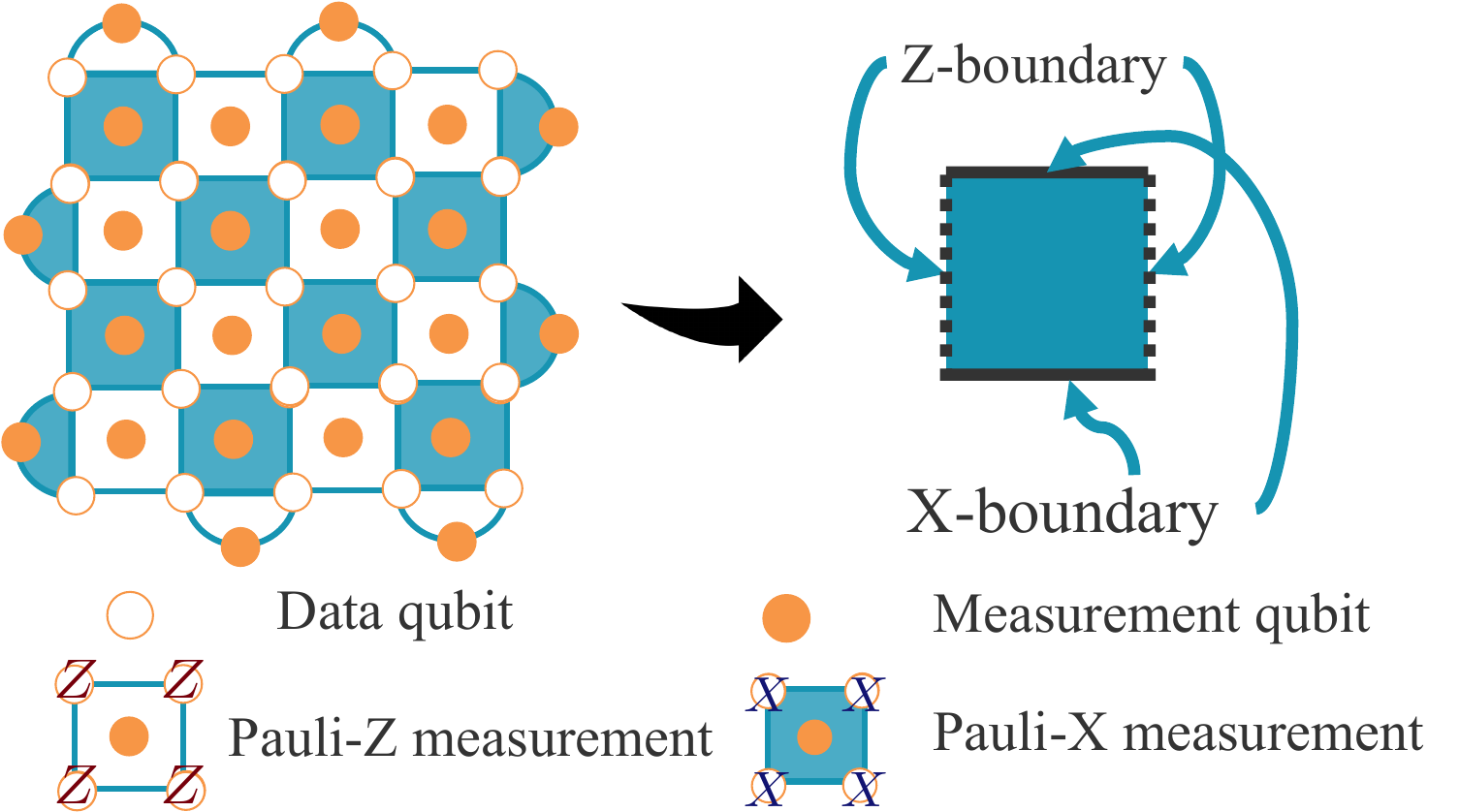}
    \caption{Illustration of a surface code with distance $d=5$. 
    }
    \label{fig:surface_code}
    \vspace{\figvspaceelim}
\end{figure}

To decrease the error rates of qubits, we need to integrate QEC, i.e., encoding qubits into redundant space to enable error detection and correction during the computation. 
Surface codes~\cite{kitaev2003quantum,bravyi1998quantum,Bombin2007} are known as one of the most promising quantum error-correcting codes because their error-correction performance is high and they can be implemented only with the nearest-neighboring interactions between qubits allocated on two-dimensional~(2D) grids. Thus, the current standard FTQC architectures are based on the surface codes~\cite{fu2019eqasm,byun2022xqsim,holmes2020nisq+,ueno2022qulatis,duckering2020virtualized}.
A logical qubit with surface codes is represented by physical data qubits placed on a 2D grid, as shown by the white circles in Fig.~\ref{fig:surface_code}. 
Errors on data qubits are detected by parity checks on subsets of data qubits. This process is called syndrome measurements. White and blue faces in this figure represent the pattern of syndrome measurements and correspond to two types of measurement, Pauli-$Z$ and -$X$, on the data qubits at the vertices of the face, respectively. A measurement qubit is placed for each face to enable up to four-qubit Pauli measurements with nearest-neighboring operations, which are shown as orange circles in the figure. 
During computation, errors can occur not only in the data qubits but also in the measurement qubits. Therefore, error locations are identified sequentially based on the outcomes of repeated syndrome measurements. One syndrome measurement cycle is called a code cycle, and one code cycle takes at least about $1 {\rm \mu s}$~\cite{fowler2012surface,holmes2020nisq+,ueno2021qecool}. 
The number of physical qubits along one side is called a code distance $d$, since surface codes can detect any simultaneous errors acting on less than $d$ qubits. 
In the arrangement shown in the figure, the sides on the left and right (top and bottom) are referred to as $Z$-($X$-)boundary, respectively. This is because Pauli-$Z$(-$X$) unitary operations along with the $Z$-($X$-)boundary result in a Pauli-$Z$(-$X$) unitary operation on encoded logical qubits. In the following discussion, we frequently represent a surface-code patch with a blue cell, where the solid and broken lines correspond to $X$- and $Z$-boundaries, respectively.

\subsection{Logical Operation on Surface-Code Qubit}
To perform an arbitrary quantum computation on FTQCs, a universal operation set for logical qubits is necessary. Some finite operation sets are known to be universal, i.e., they can construct an arbitrary operation on qubits~\cite{nielsen2002quantum}. We focus on the following universal set since they are popular in constructing FTQCs~\cite{fowler2018low}; three state-preparation operations (zero-state $\ket{0}$, plus-state $\ket{+}$, magic-state $\ket{A}$), five unitary operations (Pauli-$X, Y, Z$ operations, Hadamard operation $H$, phase operation $S$), and four Pauli measurement operations (one-qubit Pauli measurement $M_X, M_Z$, and two-qubit Pauli measurement $M_{XX}, M_{ZZ}$).
Among these operations, magic-state preparation, Hadamard operation, phase operation, and two-qubit Pauli measurements have non-negligible latency, and the latency of the other operations can be neglected.

Code deformation and lattice surgery are crucial concepts for enabling the above operations only with the nearest neighboring operations while maintaining fault tolerance~\cite{fowler2018low,litinski2019game}. They are realized by modifying the syndrome measurement patterns of surface codes. If two $X$- and $Z$-boundaries are separated by $d-1$ blue and white faces, respectively, they can also store one-qubit information with code distance $d$. During the computation, we can fault-tolerantly change the syndrome measurement patterns, and some changing patterns are known to effectively result in logical operations on qubits.

\begin{figure}[t]
    \centering
    \includegraphics[width=0.5\textwidth]{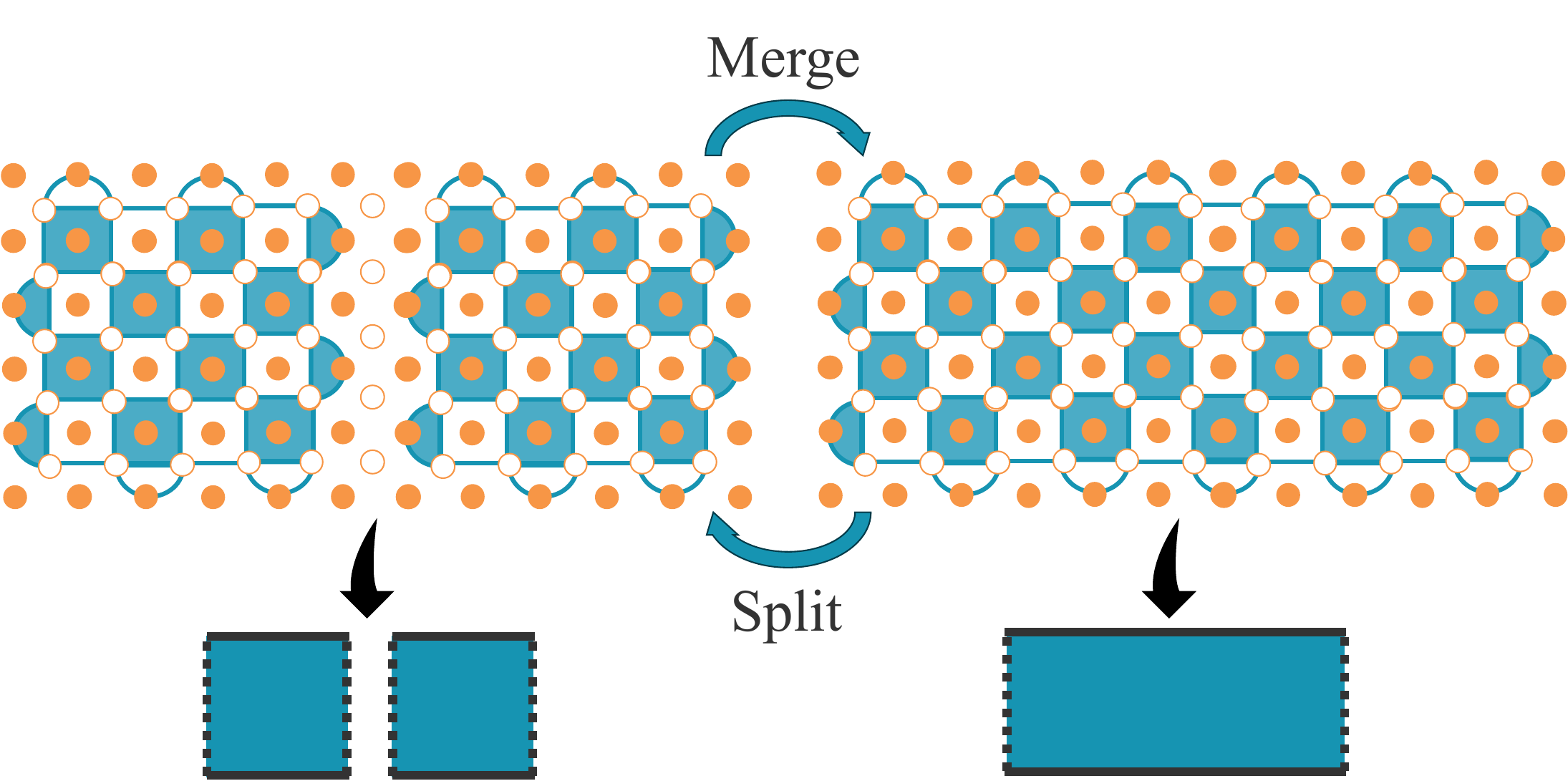}
    \caption{A procedure of lattice surgery. The merge and split process results in logical Pauli-$ZZ$ measurement.}
    \label{fig:measurement}
    \vspace{\figvspaceelim}
\end{figure}

For instance, lattice surgery can be used for performing two-body logical Pauli measurements~\cite{horsman2012surface, Vuillot_2019}. 
Suppose that two surface-code qubits are independently error-corrected. At the beginning of performing lattice surgery, a new set of parity-check faces are added using qubits between the cells to connect the two $Z$-($X$-)boundaries. After repeating the new set of syndrome measurements for a while, the code returns to the original set. This procedure effectively acts on two logical qubits as logical Pauli-$ZZ$(-$XX$) measurements. Fig.\,\ref{fig:measurement} illustrates the procedure of lattice surgery with physical and simplified views. If there is a path of unused qubits, we can perform lattice surgery operations in the same manner.
To maintain the code distance $d$ against errors, $d$ repeats of syndrome measurements are necessary for reliable error estimation after changing syndrome measurement patterns. Thus, it is convenient to define a $d$ code-cycle period as one code beat, which is used as the basic unit of computing time. In practical applications, the code distance is expected to be between 11 and 31, so one code beat is approximately $10 {\rm \mu s}$ to $50 {\rm \mu s}$. The latency of lattice surgery is one code beat.

\begin{figure}[t]
    \centering
    \includegraphics[width=0.5\textwidth]{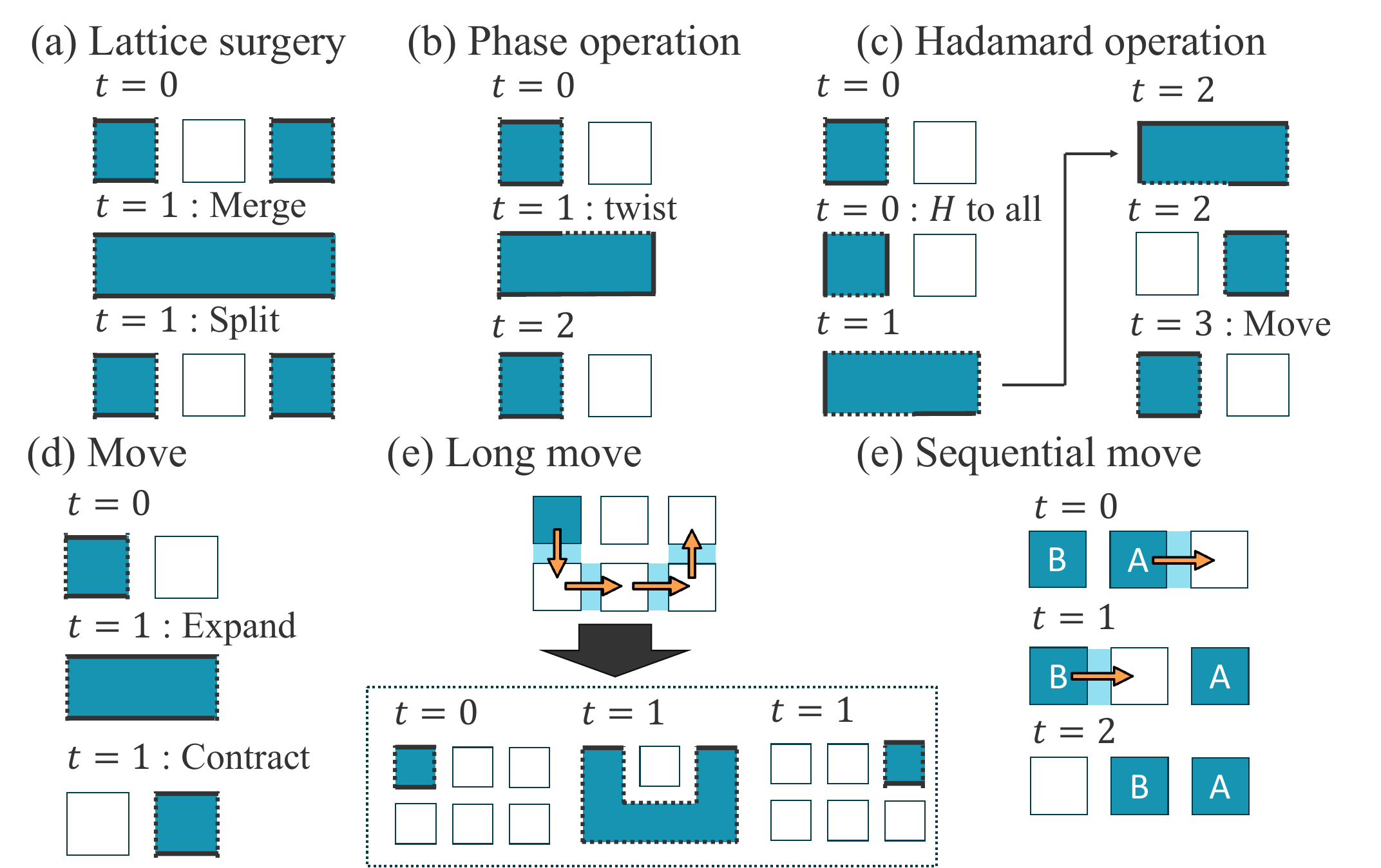}
    \label{fig:logical_operation}
    \caption{(a, b, c) Summary of logical operations. (d, e, f) Move operations. $t$ represents the elapsed code beats.}\label{fig:operation_and_move}
    \vspace{\figvspaceelim}
\end{figure}

Logical $H$ and $S$ operations can also be realized similarly by switching syndrome measurement patterns several times~\cite{fowler2018low,brown2017poking,beverland2022assessing,litinski2019game}. $H$ and $S$ operations temporally demand an additional surface-code cell and take three and two code beats, respectively. The preparation of magic states is performed via a specialized procedure called magic-state distillation~\cite{bravyi2005universal,haah2018codes,bravyi2012magic} and is known to be one of the bottlenecks in FTQC. The generation of magic states is typically performed in a dedicated space called magic-state factories (MSFs).
Code deformation can be used for expanding and reducing the patch size of logical qubits without changing the state of logical qubits, and we can move the position of a logical qubit by combining them if there is a path~\cite{litinski2019game,fowler2018low}.
The time and process required for these primitive surface-code operations are listed in Fig.~\ref{fig:operation_and_move}.
A notable feature of FTQC based on this strategy is the need for auxiliary surface-code cells to perform primitive FTQC operations.

\subsection{Major Applications and Components}
\label{subsec:major_applications}
Since the advantage of quantum computers is supported by complexity-theoretic speed-up, the evaluation and optimization for FTQC architectures should be focused on typical and bottleneck structures of quantum algorithms for various application targets rather than general  programs~\cite{babbush2021focus,beverland2022assessing}.
Major FTQC applications can fall into two classes. The first class is the analysis of quantum materials, where material structures are typically provided as a matrix called Hamiltonian represented by a linear combination of Pauli operators, i.e., $H = \sum_{i=0}^{L-1} \alpha_i P_i$ with positive coefficients $\alpha_i$ and Pauli operators $P_i$. By using a set of techniques known as quantum signal processing and linear combination of unitaries~\cite{low2017optimal}, most analysis tasks for Hamiltonian, such as calculating ground energy, simulating dynamics, and other transformations, can be performed by two unitary operations with the following actions~\cite{martyn2021grand}.
\begin{itemize}
\item SELECT: $U_S (\sum_{i=0}^{L-1} \ket{i}\ket{\psi_i}) = \sum_{i=0}^{L-1} \ket{i} (P_i \ket{\psi_i})$
\item PREPARE: $U_P \ket{0} = \sum_{i=0}^{L-1} \sqrt{\frac{\alpha_i}{\sum_j{\alpha_j}}} \ket{i}$
\end{itemize}
Since these components can be implemented with fewer magic states compared to other approaches such as Trotter-expansions~\cite{yoshioka2023hunting}, the existing resource estimation and FTQC compilation focus on efficient synthesis and scheduling of the above two operations~\cite{babbush2018encoding,lee2021even,yoshioka2023hunting,beverland2022assessing}. The ratio of execution times for these components is calculated in Fig.S11 of Ref.\,\cite{yoshioka2023hunting}, and SELECT would dominate more than 80{\%} runtime of the whole execution when considering instances of condensed matter Hamiltonian with quantum advantages. 

\begin{figure}[t]
    \centering
    \includegraphics[width=0.50\textwidth]{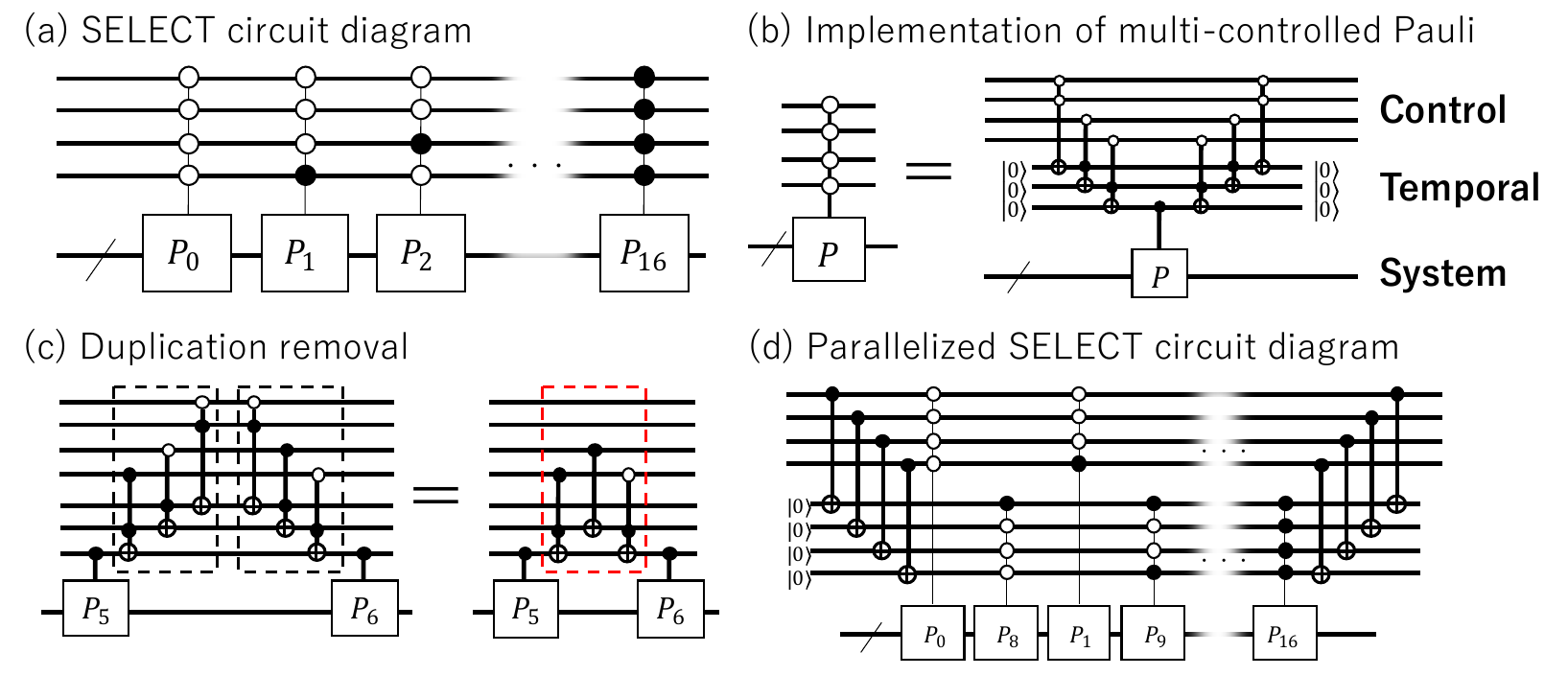}
    \caption{Circuit diagrams and optimizations of SELECT.}\label{fig:select_circuit}
    \vspace{\figvspaceelim}
\end{figure}

A naive implementation of the SELECT operation can be constructed as an iteration of multi-qubit-controlled Pauli gate $T_i = \ket{i}\bra{i} \otimes P_i + (I-\ket{i}\bra{i})\otimes I$ for $0\leq i < L$, as shown in Fig.\,\ref{fig:select_circuit}a.
Each $T_i$ can be implemented with a ladder of Toffoli gates, and qubits are classified into three registers: control, temporal, and system (Fig.\,\ref{fig:select_circuit}b).
The count and depth of Toffoli gates can be further reduced. 
Babbush~{\it et al.}~\cite{babbush2018encoding} showed that some Toffoli gates are canceled out, as shown in Fig.\,\ref{fig:select_circuit}c. Also, each Toffoli gate in SELECT operations can be decomposed into fewer $T$-gates than general cases. 
Yosihoka~{\it et al.}~\cite{yoshioka2023hunting} and Boyd~\cite{boyd2023low} showed that the depth of Toffoli gates can be reduced by creating copies of $k$ control registers, i.e., $\sum_{i=0}^{L-1} \ket{i}\ket{\psi_i} \mapsto \sum_{i=0}^{L-1} \ket{i}^{\otimes k} \ket{\psi_i}$, and applying Toffoli gates in parallel (Fig.\,\ref{fig:select_circuit}d). While further fine-tunes are available, see these references for more detailed optimizations~\cite{babbush2018encoding,yoshioka2023hunting,boyd2023low}. 

The other and longer-term application fields are solving integer problems, such as factoring, where problems are provided as integers or equations. A major difficulty in implementing this class of applications is implementing non-Clifford integer-arithmetic circuits, such as integer addition and multiplication. In Sec.\,\ref{subsec:motivation_memory_pattern}, we focus on typical examples of these two dominant application fields, SELECT and integer-multiply circuits.

\section{Motivation}\label{sec:motivation}
Since logical qubit operations require additional empty space, strategies for mapping and control of logical qubits affect the performance and memory density of FTQC architectures. 
As explained later, all the existing strategies request that all the logical operations can be executed within constant code beats at the cost of low memory density.
The purpose of this paper is to show high memory-density architecture based on load/store architecture, where the positions of logical qubits are managed by high-memory-density regions and computation is performed at a dedicated small space. While this choice imposes a penalty of longer and variable latency for memory access, we numerically show that these latencies would be highly likely concealed by other bottleneck factors in popular quantum applications.
In this section, we show evidence that the above direction is reasonable; we review the existing floorplans and mapping strategies in Sec.\,\ref{subsec:motivation_floorplan} and show that static analysis on benchmark programs indicates that fast random access is not demanded in practice in Sec.\,\ref{subsec:motivation_memory_pattern}.

\subsection{Existing Floorplan Strategies}
\label{subsec:motivation_floorplan}
\begin{figure}[t]
    \centering
    \includegraphics[width=0.3\textwidth]{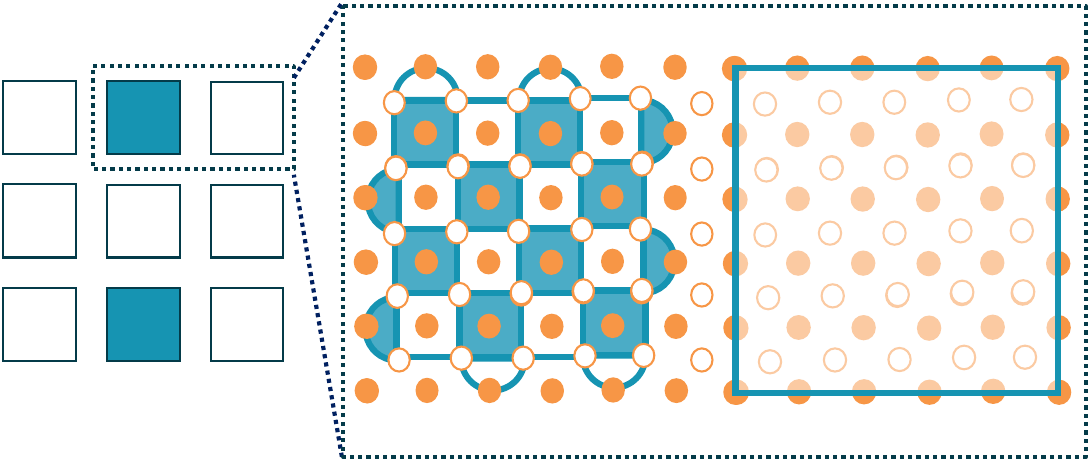}
    \caption{Diagram of the abstract surface-code in 2D grids. Blue and white cells represent data and auxiliary cells, respectively.}\label{fig:cell}
    \vspace{\figvspaceelim}
\end{figure}

When using multiple logical qubits, it is essential to determine where on the chip to allocate logical qubits auxiliary areas for logical operations. 
Suppose that the entire set of qubits is divided into surface-code-shaped cells arranged in a 2D grid as shown in Fig.~\ref{fig:cell}, and each cell is used for data cells to store qubit information or as auxiliary cells to assist logical operations. Optimization of this arrangement is equal to finding efficient floorplans, i.e., finding allocations of data and auxiliary cells to maximize the performance of FTQCs. The exploration of efficient floorplans is crucial for solving large problems with a shorter execution time and fewer qubits. 

\begin{figure}[t]
    \centering
    \includegraphics[width=0.5\textwidth]{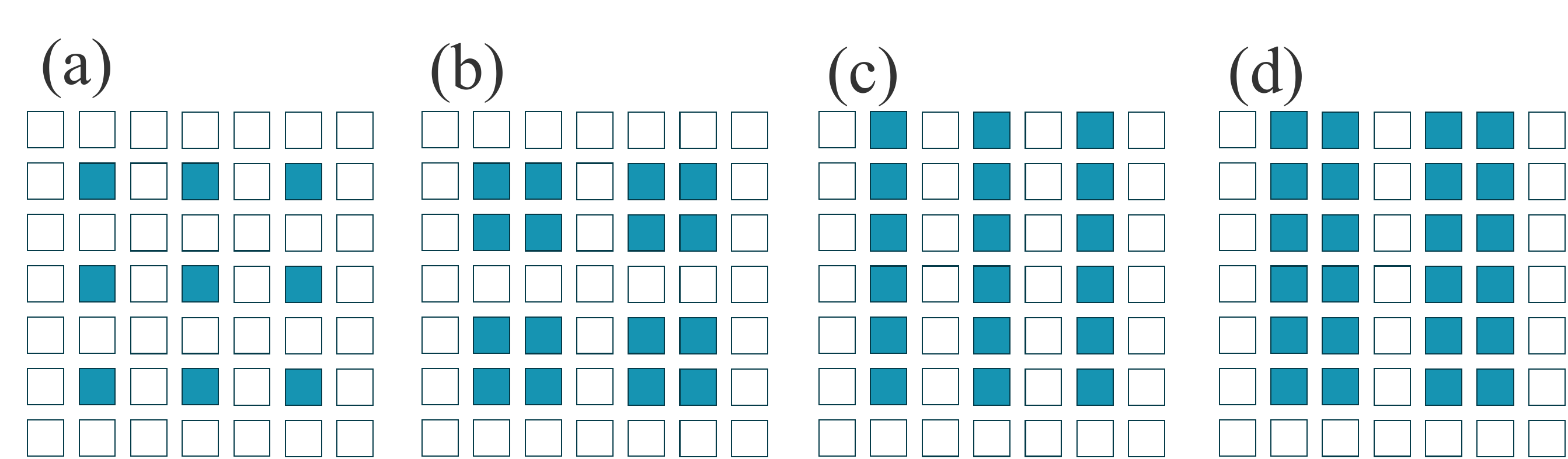}
    \caption{Floorplans proposed in the existing works. (a) 1/4-filling floorplan~\cite{beverland2022surface}. (b) 4/9-filling floorplan~\cite{chamberland2022universal}. (c) 1/2-filling floorplan~\cite{beverland2022assessing}. (d) 2/3-filling floorplan~\cite{lee2021even}.}\label{fig:conv_layout}
    \vspace{\figvspaceelim}
\end{figure}

Several floorplan patterns have been investigated in the existing works~\cite{chamberland2022universal,beverland2022assessing,lee2021even,fowler2018low,litinski2019game,beverland2022surface}. The most popular ones are listed in Fig.~\ref{fig:conv_layout}. In these floorplans, 1/4, 4/9, 1/2, and 2/3 of the whole cells are used for data cells. A common feature of these floorplans is that all the data cells have at least one neighboring auxiliary cell, and any pair of data cells has a path of auxiliary cells between them. This feature guarantees that any logical operation on any target data cells can be executed with a constant latency.

The first two floorplans guarantee that any operation can be executed without overheads for allocating auxiliary cells. 
In the 1/2-filling floorplan, either of $X$- or $Z$-boundaries is not exposed to an empty cell and cannot perform an arbitrary lattice surgery in a single code beat. This problem can be mitigated by utilizing a compact form~\cite{litinski2019game,beverland2022assessing}, where two logical qubits are constructed by two neighboring patches, and both $X$- and $Z$-boundaries are exposed to empty-cell lines, while it induces a small penalty in separately controlling them.
The 2/3-filling floorplan cannot allow one-code-beat access to logical qubits since only one of the $X$- or $Z$-boundaries of the data cells is exposed to auxiliary cells. Nevertheless, arbitrary operations can be executed with constant code beats in this floorplan.
Note that floorplans with a higher memory density are not necessarily superior to those with a lower memory density. A lattice-surgery operation on distant cells in the 1/2 and 2/3 floorplans will consume many auxiliary cells, which block the parallel execution of logical operations. 
In summary, in the existing proposals, at least 50\% of memory cells are devoted to achieving unit-time access to the logical qubits, and at least 33\% for constant-time access.

\subsection{Long Memory-Access Latency Tolerance}
\label{subsec:motivation_memory_pattern}
As explained in the previous section, about half of the integrated qubits are currently used for auxiliary cells to assist logical operations, not data storage, to guarantee unit-time random access to logical qubits. Meanwhile, the small number of available qubits is a major problem in quantum devices. This situation naturally leads to the following question; {\it Is such fast memory access required in practice?}
To address this issue, we prepared benchmarks and analyzed major application components discussed in Sec.\,\ref{subsec:major_applications}, i.e., SELECT and multiplier circuits. Then, we found that, at least in resource-restricted scenarios, fast random access memory would not improve the execution time of typical quantum programs.

For benchmarking with SELECT, we have synthesized the circuits for a $10 \times 10$ 2D Heisenberg model. For the integer-multiplying circuits, we used the multiplying circuit with 400 logical qubits developed in QASM benchmark~\cite{li2022qasmbench} and translated it into universal sets of FTQCs.
We simulated these circuits with our FTQC simulators and analyzed the memory reference trace. Here, we assumed that magic states are instantly prepared (i.e., there are sufficiently many MSFs), and logical operations can be executed in parallel if their instruction targets do not overlap. 

\begin{figure*}[ht]
    \begin{subfigure}{.62\textwidth}
        \centering
        \includegraphics[width=0.9\textwidth]{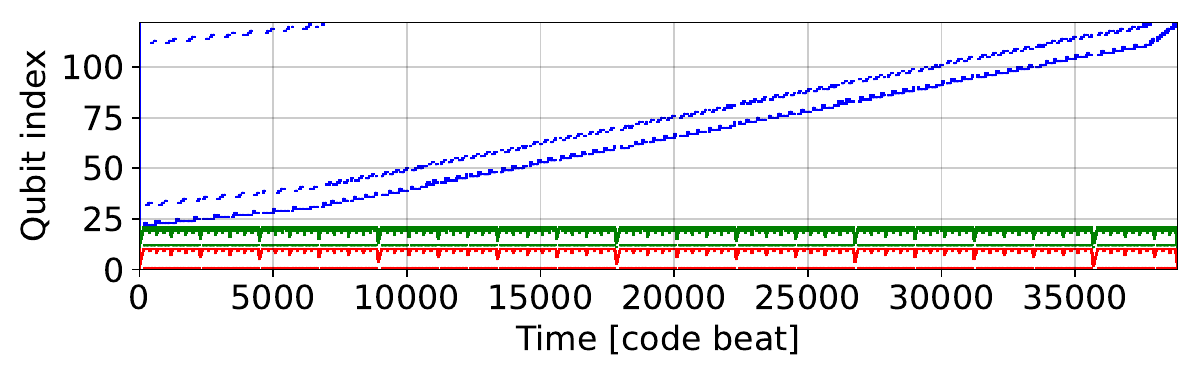}
        \caption{Memory reference timestamp for SELECT}
        \label{fig:reference_pattern_select}
    \end{subfigure}
    \begin{subfigure}{.36\textwidth}
        \centering
        \includegraphics[width=1.0\textwidth]{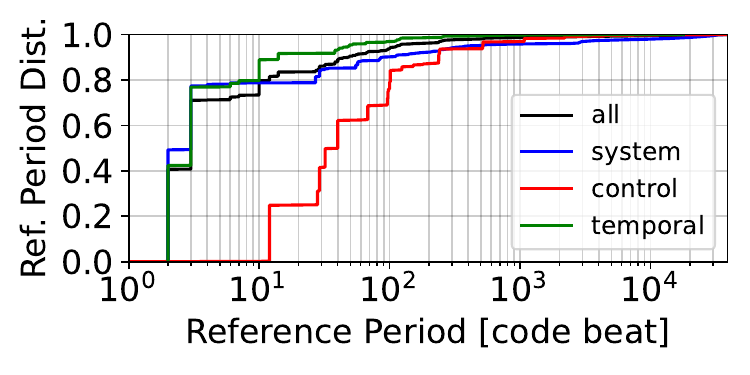}
        \caption{Reference period distribution for SELECT}
        \label{fig:reference_count_select}
    \end{subfigure}
    \begin{subfigure}{.62\textwidth}
        \centering
        \includegraphics[width=0.9\textwidth]{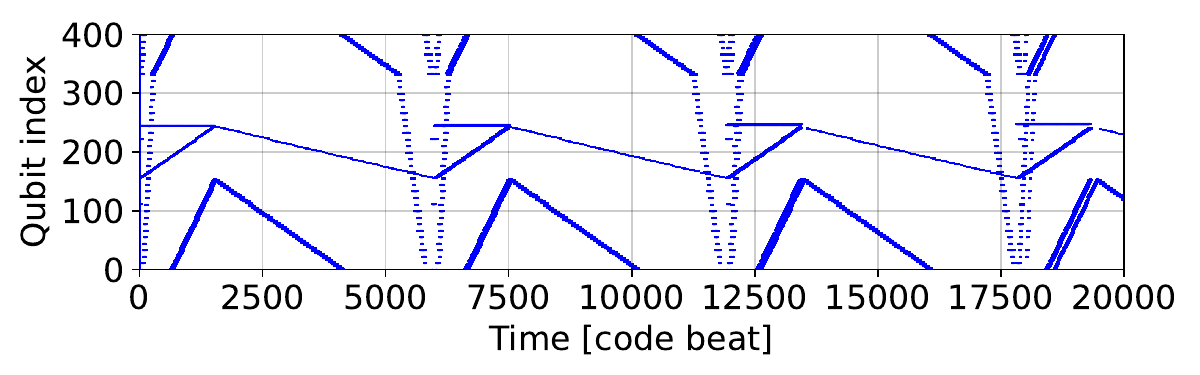}
        \caption{Memory reference timestamp for multiplier}
        \label{fig:reference_pattern_multiply}
    \end{subfigure}
    \begin{subfigure}{.36\textwidth}
        \centering
        \includegraphics[width=1.0\textwidth]{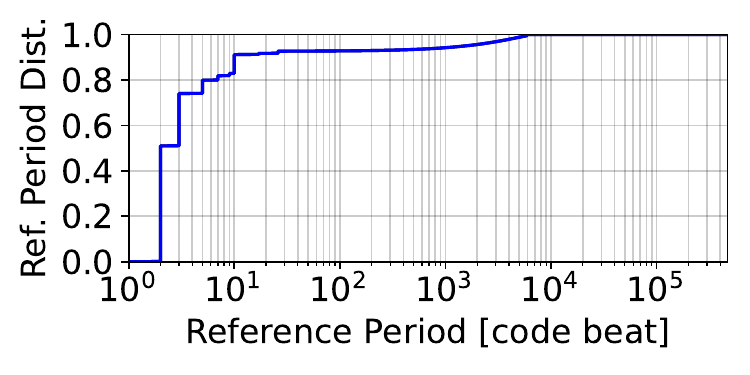}
        \caption{Reference period distribution for multiplier}
        \label{fig:reference_count_multiply}
    \end{subfigure}
        \caption{Memory reference pattern analysis for benchmark programs.}
    \vspace{\figvspaceelim}
\end{figure*}

We plotted the reference timestamp of SELECT and integer-multiplying circuits for each logical qubit in Figs.\,\ref{fig:reference_pattern_select} and \ref{fig:reference_pattern_multiply}, and we also plotted the cumulative distribution of reference periods in Figs.\,\ref{fig:reference_count_select} and \ref{fig:reference_count_multiply}, respectively. 
For the SELECT circuit, the reference timestamps and distributions of the control, system, and temporal registers are plotted in red, blue, and green, respectively. 
Since the execution time of the integer-multiplying circuit is long, we plot the first 20,000 code beats among the whole trace. 
In these traces, the SELECT and integer-multiplying circuits demand a magic state in each 11.6 and 2.14 code beats on average, respectively.

\textit{Temporal locality:} 
In both examples, cumulative distributions imply that there are many references with small periods and a few long periods between instructions. This means they have strong temporal locality, and logical qubits can be stored in space-efficient and long-latency memories during such periods with no reference for a long time. 

\textit{Spatial locality:} 
We observed clear sequential access patterns in the reference timestamps of the integer-multiplying circuits and each of the three registers of SELECT circuits. The observed patterns reflect the structure of the two programs. The SELECT operation is a sequential iteration of integer-value comparisons, and target condensed matter systems have local interactions. Typical integer-arithmetic circuits iterate bits from the lowest one to the highest one. Thus, even if the random access to the logical memory is slow, this penalty can be concealed by supporting fast sequential accesses.

\textit{Average access frequency:} 
The reference timestamps show these two benchmarks have different access frequency patterns. In the SELECT circuits, a few logical qubits in the control and temporal registers are referred to much more frequently than those in the system register. 
In contrast, the integer-multiplying circuit has almost uniform access frequencies. This implies that, in the case of SELECT circuits, the execution time can be significantly reduced by putting a small portion of logical qubits into fast-access memory regions.

\textit{Magic-state consumption:} 
Even if we utilize 176 logical cells for an MSF, one magic state is generated in every 15 code beats~\cite{litinski2019game}. Thus, both circuits demand magic states more frequently than those generated from a single MSF.
Therefore, even if fast memory access is provided, we cannot utilize it since the magic-state preparation becomes a bottleneck when a limited number of MSFs are available.

In summary, we can conclude that, in many cases and periods, unit-time random access does not contribute to reducing the execution time of popular FTQC applications in resource-restricted scenarios. In the case of SELECT circuits, most logical qubits become a target of logical instructions much less frequently than a few limited ones. Thus, fast memory access is not demanded by most logical qubits. In both examples, their references are temporally and spatially localized. Thus, there are possibilities that we can leverage these properties to make memory designs space efficient without penalty by sacrificing the access speed. Also, the consumption rates of magic states are faster than the generation rate of a single MSF. Unless massive factories are available, memory access is not a dominant bottleneck, and fast memories cannot speed up the execution time. All of these observations indicate that there is a possibility of designing more efficient FTQC architectures by exploiting these properties.
These observations might not be surprising since typical algorithms in conventional computers have the same nature, and hierarchical memory structures are well-designed to exploit them. However, to the best of our knowledge, there is no study on aggressively leveraging the access locality distribution of FTQC algorithms. These results motivate us to explore a novel FTQC architecture direction that improves memory density by reducing auxiliary cells and allowing slow memory access.

\section{Load/Store Quantum Computer Architectures}
\label{sec:proposed}
In this paper, we propose an FTQC architecture based on a novel floorplan strategy: Load/Store Quantum Computer Architectures~(LSQCA). This architecture is designed to utilize small computational regions and space-efficient memory space and support load/store instructions to move logical qubits between these two regions to leverage the observed properties of quantum programs in Sec.\,\ref{sec:motivation}. In this section, we explain the basic components of LSQCA in Sec.\,\ref{sec:proposal_overview} and provide an instruction set in Sec.\,\ref{sec:proposal_instruction}. Then, we explain a concrete design of LSQCA for surface-code-based FTQCs in Sec.\,\ref{sec:proposal_sam}. This section focuses on a simple construction for readability, and the optimization of LSQCA designs is discussed in Sec.\,\ref{sec:optimization}.

\subsection{Architecture Overview}
\label{sec:proposal_overview}
\begin{figure}[t]
    \centering
    \includegraphics[width=0.3\textwidth]{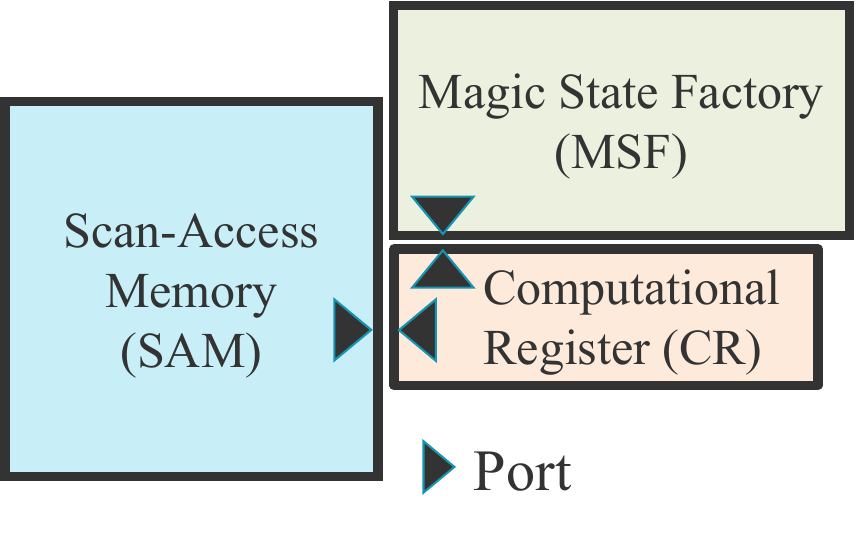}
    \caption{Conceptual diagram of the LSQCA.}
    \label{fig:LSQCA}
    \vspace{\figvspaceelim}
\end{figure}

LSQCA consists of three components: (1) Scan-Access Memory~(SAM), (2) Computational Register~(CR), and (3) Magic State Factory (MSF) for generating magic states. Each section can be arranged and connected via \textit{ports} as shown in Fig.~\ref{fig:LSQCA}.
SAM is a memory region for storing logical qubits. It consists of a large number of logical qubit cells and a small number of auxiliary cells for loading and storing operations on logical qubit cells. The detailed implementation of load/store instructions is explained in the following sections.
CR is a computing region for performing logical operations. It is filled by a small number of auxiliary cells. Logical qubits are loaded from SAM to CR, perform logical operations in CR, and then are stored in SAM.
MSFs are spaces dedicated to generating distilled magic states. The MSF is connected to CR to supply magic states when it is required. The number of MSFs can be varied to tune the performance, which is called the factory count. Since magic-state preparation can be executed in advance, we can buffer the generated magic states and conceal the latency~\cite{hirano2024magicpool}. Thus, the buffer size is another parameter for optimizing performance. The effect of these parameters on the performance is evaluated in Sec.\,\ref{sec:evaluation}.
Ports consist of surface code cells and face each other to transfer logical qubits using lattice surgery. The appropriate number of cells depends on the type of components. To communicate with MSF, we assume a single cell as a port, and we describe it later for SAM.
In addition, since the location of logical qubits changes time by time, a controller of SAM(s) is introduced, which keeps the map of variables (\texttt{M}) to the location in SAMs.

\subsection{LSQCA Instructions}
\label{sec:proposal_instruction}
The instructions of the LSQCA are listed in Table~\ref{tab:load_store}.
\begin{table*}[ht]
    \centering
    \caption{LSQCA instructions, where \texttt{LD, ST} instructions provide the abstraction for the transportation of logical qubits.}
    \begin{tabular}{|c|l|l|l|}
        \hline
        \textbf{Type} & \textbf{Syntax} & \textbf{Latency} & \textbf{Description} \\ 
        \hline
        \hline
        \multirow{2}{*}{Memory} & \texttt{LD M C} & variable & (Load) Load logical qubit from SAM to CR \\ 
        \cline{2-4}
         & \texttt{ST C M} & variable & (Store) Store logical qubit from CR to SAM \\ 
        \hline
        \multirow{3}{*}{Preparation} & \texttt{PZ.C C} & 0 beat & (Zero-Init) Initialize a logical qubit to $\ket{0}$ state \\ 
        \cline{2-4}
         & \texttt{PP.C C} & 0 beat & (Plus-Init) Initialize a logical qubit to $\ket{+}$ state \\ 
        \cline{2-4}
         & \texttt{PM C} & variable & (Magic-init) Move magic state from MSF to CR \\ 
        \hline
        \multirow{2}{*}{Unitary} & \texttt{HD.C C} & 3 beat & (Hadamard) Hadamard gate on a logical qubit\\ 
        \cline{2-4}
         & \texttt{PH.C C} & 2 beat & (Phase) Phase gate on a logical qubit\\ 
        \hline
        \multirow{4}{*}{Measurement} & \texttt{MX.C C V} & 0 beat & (Pauli-$X$ Meas) Pauli-$X$ measurement on a logical qubit and store outcome\\ 
        \cline{2-4}
         & \texttt{MZ.C C V} & 0 beat & (Pauli-$Z$ Meas) Pauli-$Z$ measurement on a logical qubit and store outcome\\ 
        \cline{2-4}
         & \texttt{MXX.C C1 C2 V} & 1 beat & (Pauli-$XX$ Meas) Pauli-$XX$ measurement on logical qubits and store outcome\\ 
        \cline{2-4}
         & \texttt{MZZ.C C1 C2 V} & 1 beat & (Pauli-$ZZ$ Meas) Pauli-$ZZ$ measurement on logical qubits and store outcome\\ 
        \hline
        \multirow{1}{*}{Control} & \texttt{SK V} & variable & (Skip) Skip next instruction if a provided value is zero\\ 
        \hline
        \hline
        \multirow{2}{*}{In-Memory Preparation} & \texttt{PZ.M M} & 0 beat & (Zero-Init) Initialize a logical qubit to $\ket{0}$ state \\ 
        \cline{2-4}
         & \texttt{PP.M M} & 0 beat & (Plus-Init) Initialize a logical qubit to $\ket{+}$ state \\ 
        \hline
        \multirow{2}{*}{In-Memory Unitary} & \texttt{HD.M M} & variable & (Hadamard) Hadamard gate on a logical qubit\\ 
        \cline{2-4}
        & \texttt{PH.M M} & variable & (Phase) Phase gate on a logical qubit\\ 
        \hline
        \multirow{4}{*}{In-Memory Measurement} & \texttt{MX.M M V} & 0 beat & (Pauli-$X$ Meas) Pauli-$X$ measurement on a logical qubit and store outcome\\ 
        \cline{2-4}
         & \texttt{MZ.M M V} & 0 beat & (Pauli-$Z$ Meas) Pauli-$Z$ measurement on a logical qubit and store outcome\\ 
         \cline{2-4}
         & \texttt{MXX.M C M V} & variable & (Pauli-$XX$ Meas) Pauli-$XX$ measurement on logical qubits and store outcome\\ 
        \cline{2-4}
         & \texttt{MZZ.M C M V} & variable & (Pauli-$ZZ$ Meas) Pauli-$ZZ$ measurement on logical qubits and store outcome\\ 
        \hline
        \hline
        \multirow{1}{*}{Optimized Unitary}& \texttt{CX M1 M2} & variable & (CNOT) CNOT gate on logical qubits\\ 
        \hline
    \end{tabular}
    \label{tab:load_store}
    \vspace{\figvspaceelim}
\end{table*}
The LSQCA instructions have three types of operands, memory qubit address (\texttt{M}), register qubit identifier (\texttt{C}), and classical value identifier (\texttt{V}). The memory address and register identifiers specify the abstracted logical cell in SAM and CR, respectively. The classical value identifier is used for measuring the final computational results or storing intermediate logical measurement outcomes and enabling adaptive executions of the following instructions.

The most characteristic instructions in the LSQCA are \texttt{LD} and \texttt{ST}, which specify the address or identifier of SAM and CR and move logical qubits between them. 
These two instructions provide the abstraction to enable LSQCA, where programmers can concentrate on the semantics of the application independently to the logical qubit allocation and the implementation of sophisticated procedures for loading and storing logical qubit cells.
They are realized by using several primitive operations to move the positions of logical-qubit cells introduced in Figs.~\ref{fig:operation_and_move}~(d, e, f).
The instructions categorized into Preparation, Unitary, and Measurement types correspond to logical operations introduced in Sec.\,\ref{sec:background} in the CR region. 
Several standard quantum processes, such as magic-state teleportation or measurement-based quantum computation, need a conditional operation, i.e., performing a logical operation if a preceding measurement value is one. The \texttt{SK} operation can be used to implement such adaptive operations.
While in-memory computation is not allowed in standard load/store architectures, several in-memory instructions are added to optimize the performance, which is explained later in Sec.\,\ref{sec:opt_in_memory}.
Additionally, we define CNOT gate instruction, which is often used in standard gate sets, as a single instruction, and it can be executed with locally optimized operations to reduce latency. Its detailed explanation is in Sec.~\ref{subsec:numerical_setting}.

Several instructions have variable latency.
Memory-type instructions (\texttt{LD}, \texttt{ST}) have variable latency, since the number of steps required to pick a certain logical cell from SAM depends on their location.
The control-type instruction (\texttt{SK}) needs to wait for the correction of the target classical value, which depends on the implementation of error estimation algorithms.
The magic-state preparation (\texttt{PM}) also has a variable latency since it has to wait for magic-state generation in MSFs when the buffer is empty.

\subsection{Design of CR and SAM}
\label{sec:proposal_sam}
In this section, we explain the detailed design of a core component in our architecture, Computational Register~(CR) and Scan-Access Memory~(SAM). We propose two types of memory-efficient implementations for SAM: point SAM and line SAM, which have different characteristics in terms of memory efficiency and access latency. Point SAM aims to maximize memory efficiency, whereas line SAM focuses on access latency improvement by exploiting the nature of logical operations on surface code and spatial locality of logical qubit references. We will explain these methods in detail and analyze their characteristics and memory efficiency. Other possible designs will be explored in Sec.\,\ref{subsec:proposal_other}.

\subsubsection{CR region}
CR consists of two columns of cells, as shown in Fig.\,\ref{fig:SAM}, and is a region for logical operations where arbitrary operations can be performed immediately.
The register cells, orange cells in the figure, are located on the right top and right bottom, which possess logical qubits loaded from SAM. 
In this work, the number of register cells is fixed to be two.
%, and the necessary bit length of the \texttt{C} identifier is one. 
The left column of CR acts as a port to connect to SAM. Fig.\,\ref{fig:SAM}a shows the most compact configuration with six cells.

Since it corresponds to the register file of microprocessors, the CR size significantly impacts CPI.
Though there is room for design space exploration in the CR size, we design CR to be as small as possible while exploiting the potential of SAM in this work.
This is to maximize the memory efficiency of surface-code-based FTQC and to figure out the distinct characteristics of the concept of LSQCA.
Sec.\,\ref{sec:opt_hybrid} will cover an extensive design space exploration.

\subsubsection{Point-SAM Architecture}
\label{sec:point_scan}
\begin{figure}[t]
    \centering
    \includegraphics[width=0.5\textwidth]{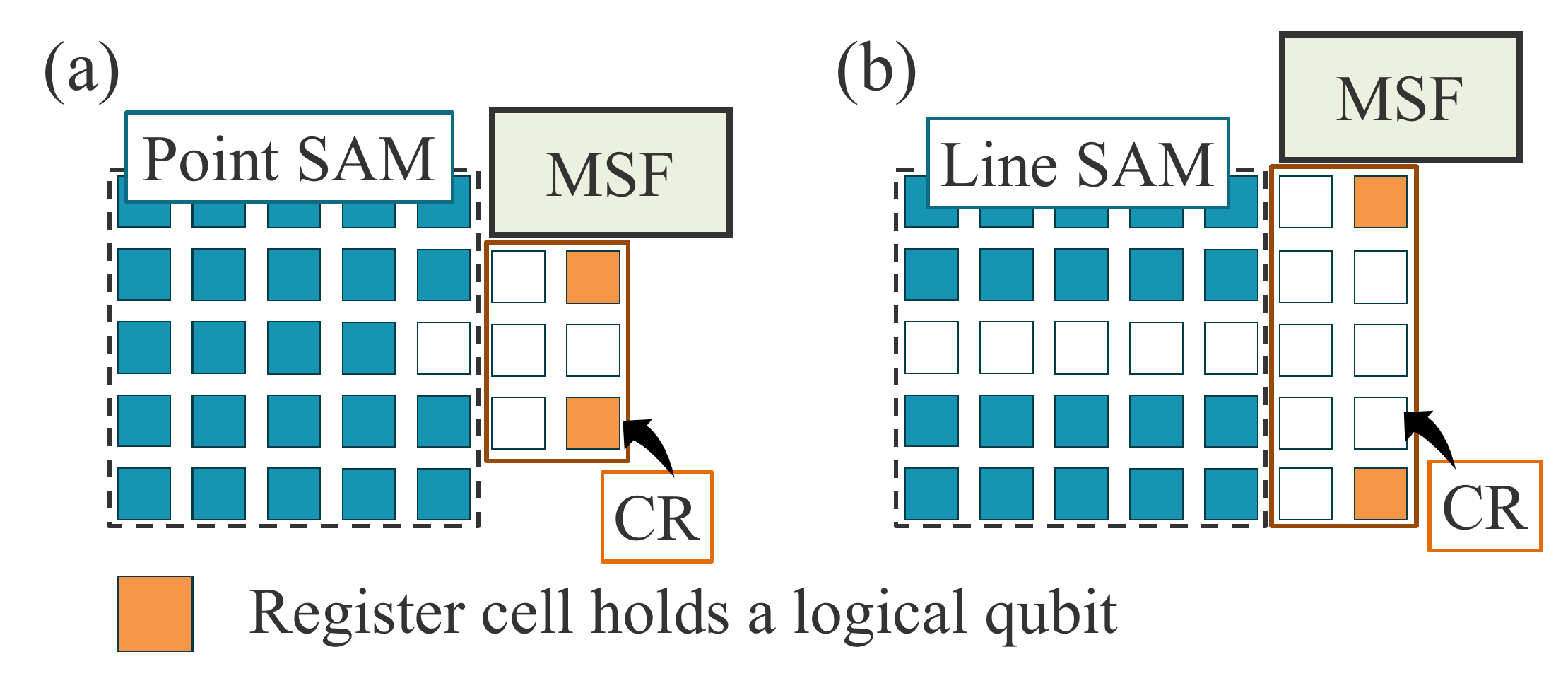}
    \caption{Schematic diagrams of SAM designs. (a) and (b) shows point-SAM and line-SAM architectures, respectively.}\label{fig:SAM}
    \vspace{\figvspaceelim}
\end{figure}

Point SAM is a design of SAM that contains only a single auxiliary cell to maximize memory efficiency.
This auxiliary cell is named a {\it scan cell} and is used for modifying the location of data cells.
Supposing that $n$ be the size of memory registers, the shape of point SAM is $\sqrt{n+1} \times \sqrt{n+1}$.\footnote{If $n+1$ is not a square number, we adjust the number by reducing the cells in the bottom line.}
In the initial state, the scan cell is set to the nearest cell to the CR region in the center line, as shown in Fig.~\ref{fig:SAM}a.
The port of point SAM consists of the cells neighboring the initial position of the scan cell, which can also possess logical qubits.

\begin{figure*}[t]
    \centering
    \includegraphics[width=1.0\textwidth]{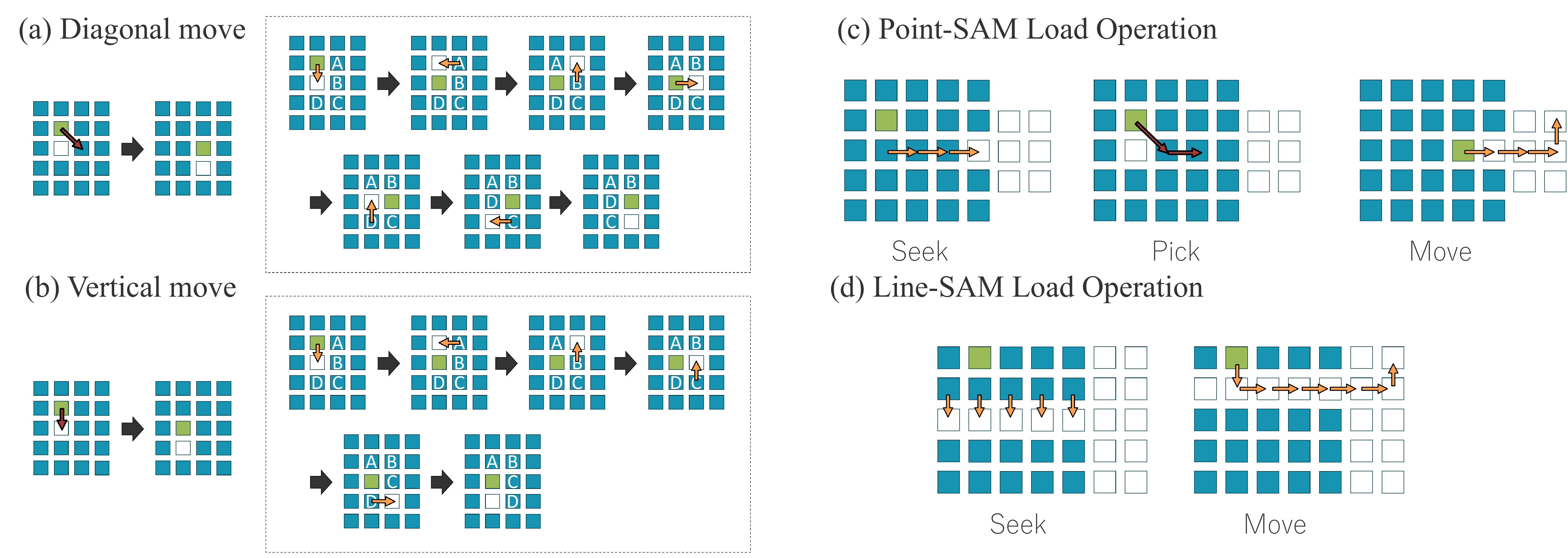}
    \caption{Procedures to execute Load operation with primitive protocols. (a) and (b) shows a way to move the target cell (green) in diagonal and vertical directions. (c) and (d) shows the load procedure for point-SAM and line-SAM architectures.}
    \label{fig:load}
    \vspace{\figvspaceelim}
\end{figure*}

The point SAM loads a target cell in a way similar to sliding puzzles. Using a set of move operations with the scan cell, the target cell can be moved to a diagonal direction with 6 beats and to a horizontal/vertical direction with 5 beats (see Figs.~\ref{fig:load}~(a, b)). 
As shown in Fig.~\ref{fig:load}c, we can load the target cell (green cell) to the CR region by repeating these moves.
First, the scan cell is moved to one of the neighboring cells of the target data cell.
Then, by repeating the move operations, we can move the position of the target cell to one of the cells of the port.
Finally, the target cell is moved to one of the cells for the logical qubits in the CR region.

Compared to the conventional floorplans, this architecture achieves asymptotically 100\% memory density but does not guarantee unit-time random access to the logical qubits. Suppose that the target cell must be moved $W$ horizontal cells and $H$ vertical cells. Then, the total steps are roughly $W+H+6{\rm min}(W, H) + 5|W-H|$ beats ignoring a small constant. This takes $7\sqrt{n}$ beats in the worst case, i.e., $W=\sqrt{n}, H=\sqrt{n}/2$. 

When one logical cell is loaded from the point SAM, there are two empty cells in the SAM region. Thus, we can use both of them to speed up the load of the second data cell. This technique enables one diagonal move per 4 beats and two vertical/horizontal moves per 6 beats. 

The store procedure for a logical-qubit cell to the original position can be performed with the reversed sequence of the load operation. On the other hand, we will utilize another strategy for store instructions to leverage the locality of the memory access. This optimization technique, locality-aware store, will be introduced in Sec.\,\ref{sec:opt_store}.

\subsubsection{Line-SAM Architecture}
\label{sec:line_scan}
Line SAM is designed with a similar concept to the point SAM, but this uses a dimensionally extended line of auxiliary cells, which is named a scan line, for loading the target logical qubits to CR.
The cell allocation of this architecture is shown in Fig.~\ref{fig:SAM}b. 
For line SAM, CR is designed to have the same height as SAM to exploit the nature of logical operations on surface code. In this design, all the cells of line SAM can be regarded as a part of a port.
The load operation of the line SAM is performed by vertically moving all the cells in the specific horizontal line that faces the scan line to move it to neighbor the line holding the target cells. Then, the target cell is transported to the CR. This process is illustrated in Fig.~\ref{fig:load}d.

The latency of the load operations in the line SAM equals the y-axis distance $H$, which is $0.5\sqrt{n}$ in the worst case. 
The store procedure can be performed with its reverse, but we will also introduce a locality-aware strategy for leveraging the locality as explained later in Sec.\,\ref{sec:opt_store}.

Compared to the point SAM, the line SAM reduces the latency by increasing scan cells.
Another feature of the line SAM is the high efficiency of continuous access to data cells on the same line without additional operations for cell movement.
In both SAM designs, the memory density becomes asymptotically close to 100\% as the size of storage $n$ increases.

\subsection{Design space and performance trade-off}
\label{subsec:proposal_other}
While we proposed the designs of CR and SAM in the last section, there are other potential designs in the LSQCA. In this section, we discuss features of SAM/CR that would affect the total performance of LSQCA and their tunability. Based on this discussion, we propose several concrete optimization methods in Sec.\,\ref{sec:optimization}.

In the proposed LSQCA architecture, the size of CR is kept small to maximize the memory density. On the other hand, this direction limits the available instruction-level parallelism~(ILP). We can extend the size of CR to allow further ILP at the cost of smaller memory density. This direction will be discussed in Sec.\,\ref{sec:opt_hybrid}.
The bandwidth of SAM can also be improved at the cost of memory density in several ways. One is increasing scan cells in SAM, which improves the average and worst-case latency of memory access at the cost of smaller memory density. In the last section, we proposed line-SAM as intermediates of the existing and point-SAM designs. Another direction is to separate SAM into several banks, which enables parallel access to memory banks and improves memory bandwidth while the memory density per SAM is reduced. This direction is discussed in Sec.\,\ref{sec:opt_multi_bank}.

The above three tunability, CR size, scan cells in SAM, and SAM bank count, trade the memory density with available ILP, memory-access latency, and memory bandwidth, respectively. How large capacity of them we need to ensure is essentially determined by the application structure. Thus, they should be designed according to the analysis of applications or adaptively tuned according to the situation. We show concrete design optimization patterns in the next section and numerically tune several optimization factors for applications in Sec.\,\ref{sec:evaluation}. We believe there are several other ways to design more sophisticated systems, but we left them as future work.

\section{Optimizations of LSQCA}
\label{sec:optimization}
While we presented a basic LSQCA in the last section, there is a huge room for improving the performance. In this section, we present four crucial ideas: multiplexing memory access via multi-bank SAM, leveraging reference locality via locality-aware store, reducing the number of load/store instructions by in-memory operations, and hybrid floorplans that use both SAM and conventional floorplans. While there are other possible optimization approaches, more extensive analysis and evaluation of other ideas are left as future work.
\begin{figure*}[t]
    \centering
    \includegraphics[width=1.0\textwidth]{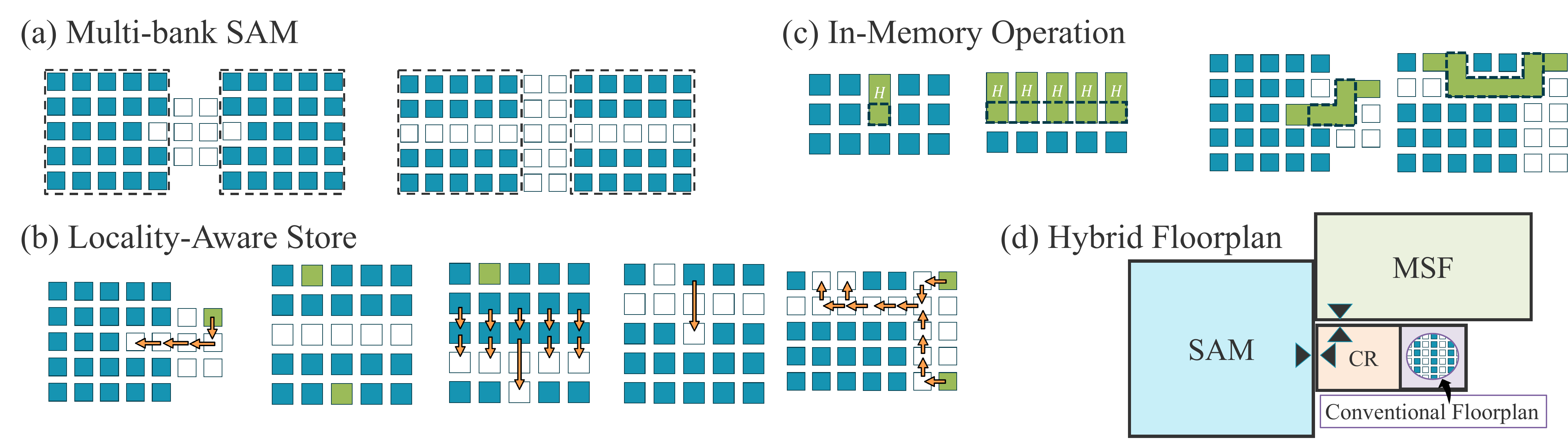}
    \caption{Optimization methods for the LSQCA.}
    \label{fig:optimization}
    \vspace{\figvspaceelim}
\end{figure*}

\subsection{Multi-Bank SAM}
\label{sec:opt_multi_bank}
In the proposed SAM designs, we can only access one target cell at a certain time. To break this limit, we can set multiple SAM regions besides the CR region and multiplex the memory access. We named an architecture with this strategy as a multi-bank SAM architecture, and each region is called a bank.
Fig.~\ref{fig:optimization}a illustrates the allocation of the 2-bank point-SAM and line-SAM architecture, respectively. This enables parallel loading from SAM banks when target data cells are stored in different banks.
While there is no essential limit to the number of banks, we need to expand the size of the CR region to let the SAM banks touch the CR. This expansion might degrade the high memory density of the LSQCA, especially in the case of the point-SAM architecture. Thus, we limit the number of banks for point-SAM architecture to two in this paper.

\subsection{Locality-Aware Store}
\label{sec:opt_store}
When the most distant data cells are repeatedly accessed, the point- and line-SAM designs impose a huge penalty in the load and store instructions. This is an unavoidable consequence of the trade-off relations between memory density and access latency when attempting to deal with arbitrary access patterns. On the other hand, conventional computers have overcome this limitation by utilizing the spatial and temporal locality inherent in most programs. Fortunately, such locality is also observed in practical quantum programs in Sec.\,\ref{sec:motivation}. Thus, we can expect that similar ideas can be employed for the LSQCA.

We can leverage temporal locality in point SAM as follows. Instead of storing the data cell in its original location, the data cells are located in an empty cell near the CR (see the left-most figure of Fig.~\ref{fig:optimization}b). This modification not only reduces the beats for store instruction but also makes the distance from the CR region to the cell small.
After executing several instructions, frequently accessed cells would typically reside closer regions to the CR, automatically forming efficient cell allocation where the access latency of recently used data becomes small. Thus, this technique can exploit the temporal locality inherent in quantum programs and can significantly reduce the average latency.

We can also modify the line SAM to leverage the spatial locality by refining the store procedure when two data cells are sequentially stored in the same bank.
If the reference patterns of quantum programs typically have spatial locality, we can expect a set of logical qubits on which a multi-qubit logical operation acts would be spatially close. Since the line SAM can quickly load the data cells in the same line, such data cells should be stored in the same or neighboring lines.
In the locality-aware store for the line SAM, when scanning multiple logical qubit cells for multi-qubit operations, the empty cells in which logical qubits are stored are moved to align with the auxiliary line. Then, two data cells are stored in the same line (see the right four figures of Fig.~\ref{fig:optimization}b).

It should be noted that this technique can be introduced without knowledge of quantum programs. Typical optimization of code deformation and lattice surgery patterns needs heavy optimization in the compilation phase, but well-tuned scheduling for a specific FTQC size would lose the portability of programs and is difficult to adapt to latency fluctuations due to probabilistic processes and error-property variations~\cite{suzuki2022q3de}. Therefore, this approach has an advantage that cannot be achieved with compilation techniques.

\subsection{In-Memory Operations}
\label{sec:opt_in_memory}
The basic LSQCA assumes every target data cell must be loaded to the CR region before operations. On the other hand, in many cases, the operation can be completed without loading everything to the CR region by using the scan cell or line as auxiliary cells. Therefore, while in-memory instructions are not defined in standard load/store architecture, we define instructions for in-memory operations as a method of LSQCA optimization since this is a natural extension of surface-code-based FTQC. In-memory instructions that can accept SAM address (\texttt{M}) are listed at the bottom of Table\,\ref{tab:load_store}.

Several single-qubit logical operations (\texttt{PZ}, \texttt{PX}, \texttt{MX}, \texttt{MZ}) do not need auxiliary cells, and they can be executed without moving the scan cell. The other single-qubit logical operations need a single auxiliary cell as described in Fig.~\ref{fig:operation_and_move}. Thus, when the scan cell/line reaches the appropriate neighboring cells of the target cell, we can perform logical operations in the SAM. The effect of this technique is significant especially in the case of point SAM, because we can skip the procedures to pick the target data cells to the CR region.
Additionally, in the line-SAM architecture, it is possible to simultaneously apply $H$ and $S$ to data cells along the line using a line of auxiliary cells as shown in the left two figures of Fig.~\ref{fig:optimization}c.~\footnote{While $S$ operations are boundary-sensitive, we can adapt this by rotating cells in advance.}

We can apply the same principle to instructions acting on two data cells, i.e., lattice surgery.
As shown in the right two figures of Fig.~\ref{fig:optimization}c, when the target data has a path of auxiliary cells to CR, we can directly perform lattice surgery operations between the two target data cells.
This will remove the last and first move of the load and store procedure, respectively.

Note that in-memory operation techniques reduce the benefit of the locality-aware store since this technique will skip store instructions. 
Thus, the choice of in-memory operation or locality-aware store should be examined for each instruction. This point will be discussed in Sec.\,\ref{subsec:numerical_setting}.

\subsection{Hybrid Floorplan}
\label{sec:opt_hybrid}
Some logical qubits might be heavily accessed during the whole quantum program. In this case, even with the technique to leverage access locality, the latency of load and store instructions would not be negligible.
Also, since the ILP is limited by the number of register cells, there is a significant penalty in the execution time if it does not meet the ILP demanded by applications.

A possible solution to mitigate this problem is to expand the CR regions. To this end, instead of varying the number of cells in CR, we attach a conventional floorplan to effectively increase the number of register cells, as shown in Fig.~\ref{fig:optimization}d. 
By allocating heavily accessed logical qubits to the conventional regions, we can reduce the number of load/store instructions to facilitate faster and more memory-efficient computation and can improve the available ILP.
When we use a hybrid floorplan, the size of the computational register identifier (\texttt{C}) is extended according to the area of introduced regions with a conventional floorplan. 

While the use of the hybrid floorplan degrades the memory density, we can improve the available ILP and reduce the number of load/store instructions by tuning the size of conventional floorplans while keeping the penalty of the LSQCA modest. The sensitivity of the size of conventional floorplans to the execution time highly depends on the target quantum programs, which is examined in Sec.\,\ref{sec:result_hybrid}.

\section{Performance Evaluation}\label{sec:evaluation}
This section evaluates the performance of the LSQCA in comparison to existing strategies with realistic instances.

\subsection{Numerical Setting}\label{subsec:numerical_setting}
\textit{Performance evaluation method:} The performance of our architecture is evaluated with Code beats Per Instruction (CPI) and memory density. To this end, we developed a code-beat-accurate simulator that can track the primitive operations decomposed from load/store instructions and can calculate the execution time and CPI, which is the ratio of execution time to the command count of LSQCA. 
The bank and factory counts are varied in each benchmark. The hybrid floorplan is not used in the benchmark in Sec.\,\ref{sec:result_bench} but is used in Sec.\,\ref{sec:result_hybrid}. We always use locality-aware store when the store instructions are executed.
In the simulation and evaluation, we utilized the following assumptions. 
When we encounter conditioned operations, we always choose the taken path, i.e., we always choose the path with more instructions. 
We ignore instructions with negligible latency, such as Pauli unitary operations, in our evaluation.
Our memory density includes SAM banks and CR but excludes MSFs, as their memory density depends on their specific design. 
We use a popular design of MSFs that can generate one magic state per 15 code beats, proposed by Litinski~\cite{litinski2019game}. 
The number of magic-state buffers is set to $2n$ for a given number of factories $n$.
The shape of the SAM banks is assumed to be either $L \times L$  or  $L \times (L+1)$, and under these conditions, the configuration is set to maximize memory density. Other shapes are considered as future work.
When using multiple SAM banks, logical qubits are distributed sequentially to all the banks in order. High-performance assignment of logical qubits to multiple banks is left as future work. 
For CR, as mentioned in Sec.~\ref{sec:proposal_sam}, it assumes the smallest unit, and after logical cells are loaded and computations are executed, they are immediately stored.
Note that since we count the execution time in the unit of code beats and evaluate memory density in the unit of cells, the performance is independent of the chosen code distance or error rates of physical qubits.

\textit{Benchmark program compilation:} Each benchmark program is decomposed into the load/store instructions as follows.
First, they are represented by Clifford operations ($H$, $S$, and CNOT gates), $T$ gates, and single-qubit Pauli measurements. 
Each $T$ gate is decomposed into Pauli-$ZZ$ measurement on target qubit and magic state, and $S$ gate on target qubit according to the measurement outcome. Then, each gate is translated to our instruction set by attaching load/store instructions to each gate.
When we translate single qubit gates, in-memory instructions are always used. For two-qubit gates, we use in-memory operations for the second-operand qubit if the latency for loading the first operand qubit is faster than the other. To manage this at runtime, we defined optimized unitary instruction CNOT in Table.\,\ref{tab:load_store}. Also, we use an in-memory operation for the Pauli-$ZZ$ measurement induced from $T$ gates. The other gates are translated with attaching load and store instructions. 
We can consider further optimizations for compilation schemes, such as adaptive translation or instruction reordering, which are left as future work.

\textit{Baseline setting:} 
Since there are several floorplan strategies in the existing work, we compare the LSQCA performance with that of an optimistic baseline, which we call a conventional floorplan, designed by combining the advantages of several existing floorplans. 
We set the memory density of the conventional floorplan as 50\% from Fig.\,\ref{fig:conv_layout}c since this is the maximum density that allows random access with negligible routing overheads. 
While the ILP is limited by the memory density and qubit locations due to path conflicts of lattice surgery, we assume there are no path conflicts in the conventional floorplan. 
This means the logical qubit mapping on the conventional floorplan does not affect its performance, which makes evaluation simple. 
When we consider the hybrid floorplan in Sec.\,\ref{sec:result_hybrid}, we use conventional floorplans for the hybrid floorplan. 
The above assumption is optimistic for the baseline or equally affects the performance of the proposed architecture and baseline. 
Thus, this comparison is sufficient to show the advantage of LSQCA over the existing strategies, while this assumption makes evaluation independent from detailed optimization on the baseline, such as logical qubit mapping optimization according to the program structure.

\subsection{Evaluation with Benchmark Programs}
\label{sec:result_bench}
\begin{figure}[tbp]
    \centering 
    \includegraphics[width=1.0\linewidth]{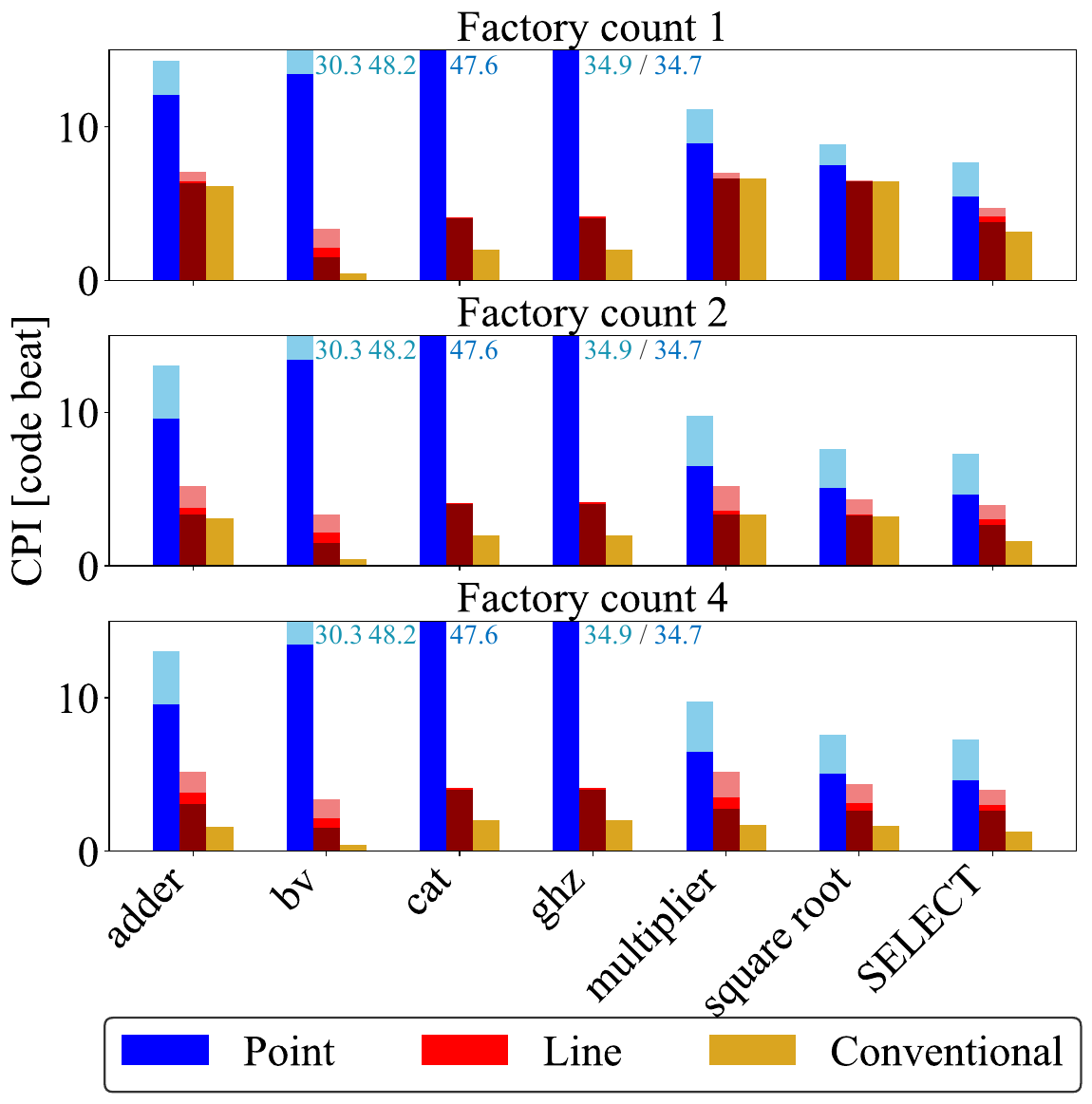}
    \caption{Performance benchmark results. Each figure shows the CPI with a different number of MSFs. 
    }\label{fig:result_execution_time}
    \vspace{\figvspaceelim}
\end{figure}

First, we show the results of the execution time evaluation for several actual quantum programs. 
In addition to SELECT and multiplier quantum programs used in Sec.\,\ref{sec:motivation}, we selected five benchmark programs from the QASM benchmark~\cite{li2022qasmbench}: Quantum adder~(adder), Bernstein-Vazirani algorithm~(bv), preparing cat state~(cat), preparing GHZ state~(ghz), and circuits for computing the square root of a number via amplitude amplification~(square root). 
The number of required logical qubits is as follows: 433 for adder, 280 for bv, 260 for cat, 127 for ghz, 400 for multiplier, 60 for square root, and 143 for SELECT circuit for the $11 \times 11$ 2D Heisenberg model.
The results are shown in Fig.~\ref{fig:result_execution_time} with several factory counts. The blue, red, and yellow bars correspond to the performance of LSQCA with point SAM and line SAM and that with conventional floorplans, respectively. The number of banks is also tested with settings of 1 and 2 for point SAM, and 1, 2, and 4 for line SAM. The darker the color, the greater the number of banks.

The  LSQCA successfully improves memory density, which is given as the number of required logical qubits by the application over the total logical qubit count in the architecture. For instance, in the case of a multiplier with only one factory, using Line SAM achieves approximately $400/462\simeq87\%$ memory density with 6$\%$ execution time overhead. More comprehensive results appear in Fig.~\ref{fig:result_density_cpi_all}.
When the number of factories is limited to one, as assumed in limited-scale FTQC, the comparison between LSQCA and conventional floorplans shows significant differences for bv, cat, and ghz instances, and small differences for adder, multiplier, square root, and SELECT instances. This is a natural consequence because the formers do not demand any magic states, and all gates can be executed without any magic-state generation delay with high parallelism. In this case, the latency of load/store operations is not concealed, resulting in a large overhead in execution time.
On the other hand, instructions in the latter circuits demand many magic states and have relatively low parallelism. In such circuits, the penalty of LSQCA is effectively concealed. The use of multiple banks enhances this trend. We note that penalties in the former benchmark (bv, cat, and ghz) are not a serious problem since they are not typically the bottleneck of typical useful quantum algorithms, while the latter examples are heavily repeated bottlenecks in practical algorithms as mentioned in Sec.~\ref{subsec:major_applications}.

As the number of MSFs increases, the bottleneck caused by the lack of magic states is alleviated. In such a regime, while the execution time of LSQCA and conventional floorplans improves, the discrepancy in the performance also expands. On the other hand, using the multi-bank technique improves the parallelism of LSQCA and can close the performance gap between LSQCA and conventional floorplans in benchmark cases with high parallelism.

Note that when we choose a configuration with a large execution time overhead, the total logical error rate increases, and code distances must be increased, which degrades the memory density improvement. Nevertheless, as far as the execution time penalty is modest, this effect would be negligible in practice. Also, the execution time overhead can be tuned with hybrid floorplan optimization introduced in the next section.

\subsection{Optimization with hybrid floorplan}
\label{sec:result_hybrid}
We perform further optimization using a hybrid layout using the same benchmark programs as in Sec.~\ref{sec:result_bench} and present the results showing the relationship between memory density and execution time. 
To tune the size of the conventional floorplan in the hybrid layout, we define a ratio of conventional floorplan $f$. Supposing the number of data cells is $n$, the size of the conventional floorplan is $nf$, and that of SAM is $n(1-f)$. We put the most frequently accessed $nf$ data cells into the conventional floorplan and the other into the SAM.

\begin{figure}[tbp]
    \centering 
    \includegraphics[width=1.0\linewidth]{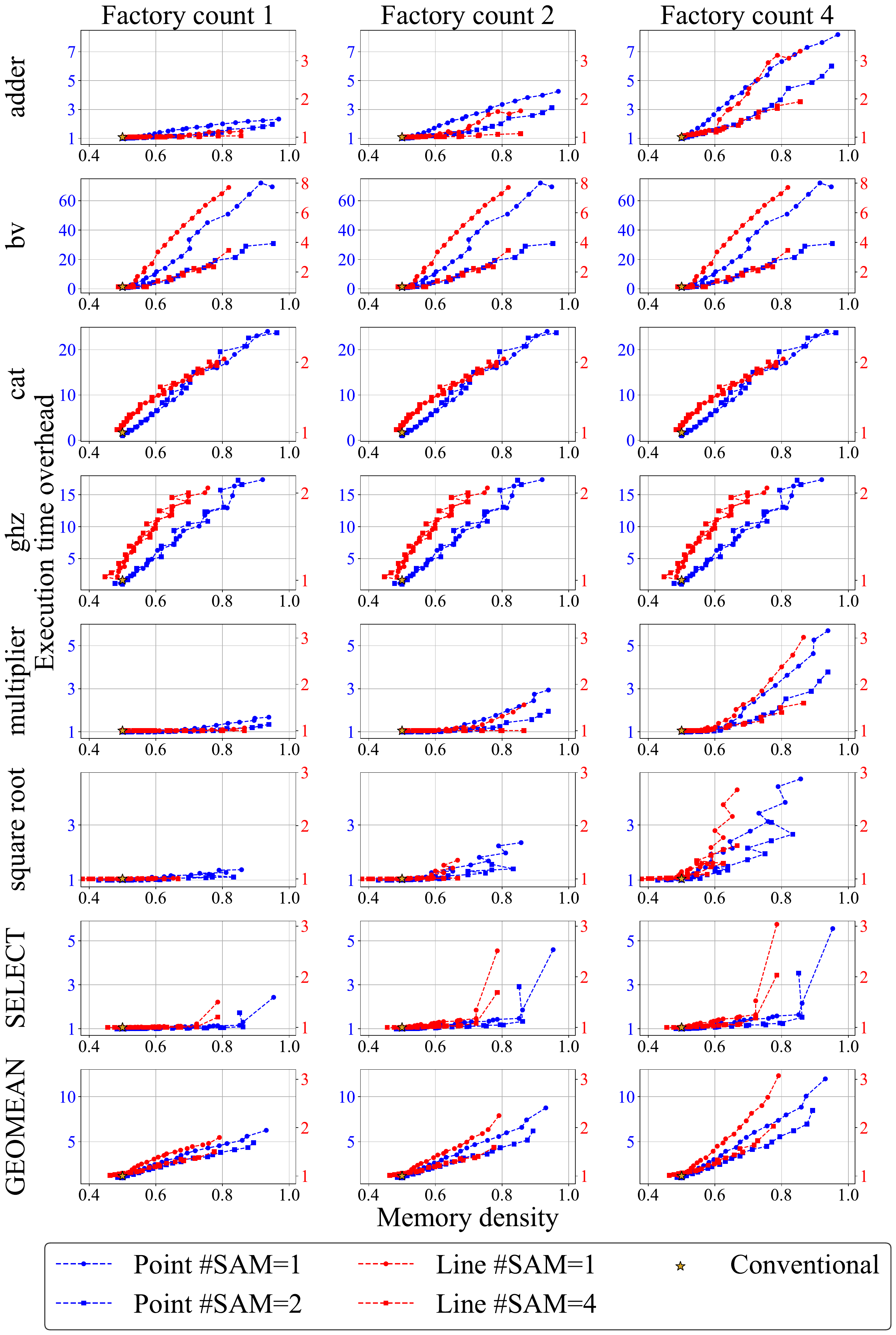}
    \caption{Performance of LSQCA with hybrid layout. Each figure shows the trade-off between memory density and execution time overhead compared to the conventional floorplan.}\label{fig:result_density_cpi_all}
    \vspace{\figvspaceelim}
\end{figure}

\begin{figure*}[ht]
    \centering
    \includegraphics[width=1.0\textwidth]{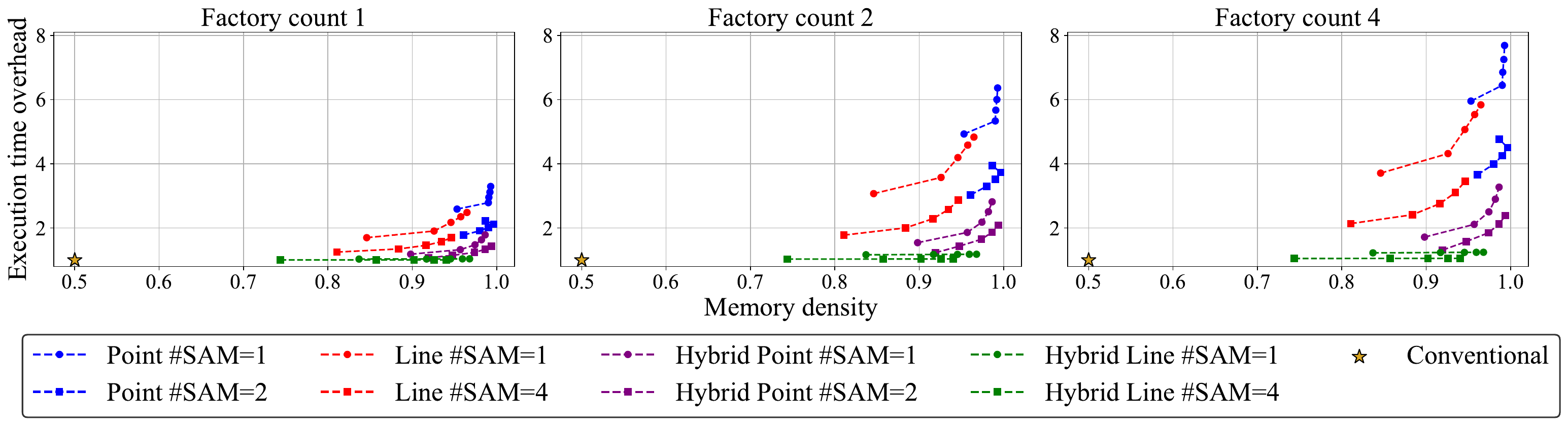}
    \caption{Performance of the LSQCA for several instance sizes of the SELECT circuits. Each figure shows the trade-off between memory density and execution time overhead compared to the conventional floorplan.}
    \label{fig:hybrid_SELECT}
    \vspace{\figvspaceelim}
\end{figure*}

The results are summarized in Fig.~\ref{fig:result_density_cpi_all}. In addition to the benchmark results, their GEOMEAN is also shown to evaluate the average performance of LSQCA. The horizontal axis represents memory density, and the vertical axis represents the overhead in execution time compared to the conventional floorplan. For each SAM layout, the ratio $f$ is varied from 0 to 1 by 0.05, and these points are connected with dashed lines. Note that the two endpoints $f=0$ and $f=1$ correspond to LSQCA without a hybrid floorplan and the baseline in the previous evaluation, respectively.

In every case, we observed the trade-off relation between memory density and execution time overhead. 
Compared to the cases of bv, cat, and ghz circuits, the execution time overhead by the hybrid floorplan is modest in the other cases, i.e., adder, multiplier, square root, and SELECT circuits. 
In the case of the multiplier, square root, and SELECT, the trend of the plots changes at a specific ratio $f$. In particular, the SELECT circuits show the most drastic trend change at $f=0.95$. We expect this is because the control and temporal registers are heavily referenced by many instructions compared to the system register, as observed in Sec.\,\ref{sec:motivation}.

Finally, we evaluated the efficacy of hybrid layouts in terms of instance sizes with the SELECT circuits. We assume the hybrid floorplan in which the data cells for control and temporal registers are placed in the conventional floorplan. The chosen instance sizes (the width of spin lattices of 2D Heisenberg models) are 21, 41, 61, 81, and 101. Their required data cells are 467, 1,711, 3,753, 6,595, and 10,235, respectively. Fig.\,\ref{fig:hybrid_SELECT} shows the evaluation results. 
The memory density increases as the instance size increases since the CR size becomes negligible. Thus, points are plotted from left to right as the instance size increases.
In resource-limited cases (instance size of 21, requiring 467 logical qubits, and with one MSF), Hybrid Point achieves approximately 92{\%} memory density with 7\% execution time overhead. In resource-abundant cases (instance size of 101, requiring 10,235 logical qubits, and with four MSFs), Hybrid Line achieves approximately 94{\%} memory density with a 6\% execution time overhead.
This means that the LSQCA reduces the qubit count for SELECT circuits with a modest penalty not only for small-size instances but also for large ones by introducing hybrid floorplans. This indicates that, when treating large circuits with heavily accessed logical qubits, LSQCA would be an indispensable concept that can significantly improve the computational capability of FTQCs.

\section{Applicable scope and related works}
\subsection{Applicable scope of LSQCA}
The essential concept of LSQCA is to divide the qubit space into computing space~(CR) and high-memory-density space~(SAM) at a logical level and improve the memory density while concealing CPI overheads; thus, the concept can be applied to any qubit device, QEC codes, and qubit topology, as far as they are error-corrected and have a logical-level access locality. 
As a practical and widely applicable design instance of the LSQCA, this paper proposed SAMs and floorplans. This design space is available in architectures with a 2D-grid array of surface-code patches with nearest-neighboring interactions, which is a standard model in many FTQC theories and architectures~\cite{fowler2018low,beverland2022assessing,chamberland2022universal,litinski2019game}. Thus, quantum devices, such as superconducting circuits~\cite{google2023suppressing,acharya2024quantum,chamberland2020building}, trapped ions~\cite{erhard2021entangling,brown2016co,monroe2014large}, and optical photons~\cite{bourassa2021blueprint,bombin2021interleaving} aim to achieve the same model at a logical level.
While some architectures can utilize more flexible connectivity by 3D integration~\cite{rosenberg20173d}, multi-chip modules~\cite{smith2022scaling}, multi-mode resonator array~\cite{duckering2020virtualized}, 2D loop-array~\cite{cai2023looped}, or optical tweezer arrays~\cite{bluvstein2024logical}, they do not allow constant-time random-access to two-logical-qubit operations unless sacrificing the memory density. For example, neutral atoms can directly move positions, but their shuttling latency increases with physical distances, while lattice surgery can be executed with a constant latency. Thus, while we might need to refurbish the concept for point- and line-SAM, the idea of LSQCA would be versatile to improve memory density by modestly sacrificing CPI in various architectures.

\subsection{Related work}
Several existing works successfully improved FTQC architectures by mixing different designs on floorplans. The major existing strategies are to apply different optimization schemes to resource-state generation. 
Holmes~{\it et al.}~\cite{Holmes2019resource} proposed a distributed design of MSFs. They extracted algorithm properties such as $T$-count and magic-state consumption rates from realistic application algorithms and utilized them to maximize the performance of FTQCs. 
Litinski~\cite{litinski2019game,litinski2019magic} extensively explored optimized floorplans dedicated to MSFs. Several hand-optimized designs are shown under trade-offs between size and throughput according to the precise analysis of the magic-state fidelities during the distillation process.
Stein~{\it et al.}~\cite{Stein2023hetarch} proposed a framework that enables the exploration of heterogeneous designs in broad layers of FTQCs. They demonstrated the flexibility of the framework in the case of superconducting qubits and represented high-level protocols such as error correction, entanglement distillation, and magic-state distillation. Using this framework and leveraging the optimized choice of low-level quantum hardware, they improved the logical error rates for QEC and resource generation protocols.
Xu~{\it et al.}~\cite{Xu2024constant} and Bravyi~{\it et al.}~\cite{Bravyi2024high} proposed an idea of using two QEC codes with different properties. The quantum low-density parity check~(qLDPC) codes are known to have good encoding rates but are difficult to execute a universal set of logical operations. They proposed qLDPC codes compatible with superconducting circuits and neutral atoms and a method to move logical qubits between qLDPC and surface code regions. Compared to these studies, the significant points of our proposal are as follows.

\textit{Breaking the existing density limit:} 
The memory density of surface-code-based architectures particular about constant-time access has a limitation to improve. Due to this reason, the proposed floorplan that can be used for general quantum programs is limited to 1/2, or 2/3 with execution overheads. The LSQCA enables the exploration of designs beyond the wall by sacrificing the constant-time access to the logical qubits, based on the observation that such fast access is not demanded in bottleneck components of major applications.

\textit{Exploiting application characteristics beyond $T$-gate:} 
Several existing works have explored the direction of optimizing the FTQC design according to the algorithm structure. Holmes~{\it et al.}~\cite{Holmes2019resource} exploit the information of magic states and $T$-gates, such as the count and consumption rate of magic states, and Litinski~\cite{litinski2019game,litinski2019magic} also exploits the detailed structures and trade-offs in MSF designs. Stein~{\it et al.}~\cite{Stein2023hetarch} proposed the framework to exploit the design space in low-level hardware, such as a heterogeneous combination of superconducting devices. 
Our paper focuses on exploiting inherent algorithm structures such as access locality and instruction parallelism in addition to $T$-gate information. Thus, we leveraged high-level information compared to Ref.\,\cite{Stein2023hetarch} and more detailed properties of algorithms compared to Refs.\,\cite{litinski2019game,litinski2019magic,Holmes2019resource}.
Analyzing these properties and leveraging them for concealing memory access latencies are first achieved in this paper and represent one of the major contributions. 

\textit{Program Portability:} 
To the best of our knowledge, most existing papers that consider qubit connectivity treat it by using quantum gates on target qubits with explicit physical locations. This means compiled object codes can be executed only for a specific target device, or connectivity needs to be resolved at runtime. While Stein~{\it et al.}~\cite{Stein2023hetarch} also studied abstracted memory, their concepts are motivated by extending design space from choices of multiple devices to high-level FTQC subroutines. Since our paper targets logical-level memory designs and latencies for general computation, the aim of abstraction is different from the existing works. Our paper separates issues of resolving logical-qubit positions from operational instructions and delegates them to abstract SAMs, which enables instruction sets independent of qubit locations. With this abstraction, this paper first provides application-level portability, i.e., common object codes can be executed in every instance of LSQCA in an efficient manner.

Since the LSQCA targets the trade-off between the memory density and CPIs for a given MSF configuration, our proposal is compatible with ideas to optimize MSFs.
We also note that while magic-state counts and MSFs have been considered dominant bottlenecks in FTQCs, recent theoretical development significantly reduced the cost of them~\cite{low2017optimal,litinski2019game,babbush2018encoding,lee2021even,beverland2022assessing,lin2022heisenberg,litinski2019magic,akahoshi2024partially,gidney2024magic,gidney2023tetrationally}.
Thus, the design space exploration of FTQCs focusing not only on MSFs and magic-state counts like this proposal would become more vital in the future.

\section{Conclusion}
\label{sec:conclusion}
In this paper, we propose LSQCA, designed to implement a Load/Store architecture on FTQC. This architecture enables efficient storage for rarely-accessed qubits and virtualized addressing for abstract components, SAM banks, CR, and MSFs.
We proposed two designs that can be implemented with primitive operations for surface-code-based FTQCs: the point-SAM architecture and the line-SAM architecture. 
Our numerical evaluation with practical instances indicates that our architecture improves memory density while concealing the penalty of memory access latency, which implies that our architecture is useful for accelerating the demonstration of quantum advantages with FTQCs.

\section*{Acknowledgements}
This work was supported by JST PRESTO Grant Number JPMJPR2015, JST Moonshot R\&D Grant Number JPMJMS2061, JPMJMS2067, and JPMJMS226C, JST CREST Grant Number JPMJCR23I4 and JPMJCR24I4, MEXT Q-LEAP Grant Number JPMXS0120319794 and JPMXS0118068682, the Center of Innovation for Sustainable Quantum AI, JST Grant Number JPMJPF2221, JST SPRING, Grant Number JPMJSP2108, JSPS KAKENHI Grant Numbers 22H05000, 22K17868, and 24K02915, RIKEN Special Postdoctoral Researcher Program.

\newpage
%%%%%%% -- PAPER CONTENT ENDS -- %%%%%%%%

%%%%%%%%% -- BIB STYLE AND FILE -- %%%%%%%%
\bibliographystyle{IEEEtranS}
\bibliography{bibs}
%%%%%%%%%%%%%%%%%%%%%%%%%%%%%%%%%%%%

\end{document}